\documentclass[amsmath,amssymb,aps,prd,twocolumn]{revtex4-2}
\usepackage{graphicx}
\graphicspath{{figures/}}
\usepackage{dcolumn}
\usepackage{subfigure}
\usepackage{bm}
\usepackage{footnote}
\usepackage{url}
\usepackage{xspace}
\usepackage{xcolor}

\newcommand\apjl{The Astrophysical Journal Letters}
\newcommand\aj{The Astronomical Journal}
\newcommand\mnras{Monthly Notices of the Royal Astronomical Society}
\newcommand\aap{Astronomy and Astrophysics}

\newcommand\pasa{Publications of the Astronomical Society of Australia}

\begin{document}
%\begin{CJK*}{UTF8}{gbsn}

\title{Multi-messenger observations of double neutron stars in Galactic disk \\ 
with gravitational and radio waves}

\author{Wen-Fan Feng}
\affiliation{MOE Key Laboratory of Fundamental Physical Quantities Measurements, Hubei Key Laboratory of Gravitation and Quantum Physics, PGMF, Department of Astronomy and School of Physics, Huazhong University of Science and Technology, Wuhan 430074, China}

\author{Jie-Wen Chen}
\affiliation{MOE Key Laboratory of Fundamental Physical Quantities Measurements, Hubei Key Laboratory of Gravitation and Quantum Physics, PGMF, Department of Astronomy and School of Physics, Huazhong University of Science and Technology, Wuhan 430074, China}

%\author[0000-0001-8990-5700]{Yan Wang}
\author{Yan Wang}
\email{ywang12@hust.edu.cn}
\affiliation{MOE Key Laboratory of Fundamental Physical Quantities Measurements, Hubei Key Laboratory of Gravitation and Quantum Physics, PGMF, Department of Astronomy and School of Physics, Huazhong University of Science and Technology, Wuhan 430074, China} 

\author{Soumya D.~Mohanty}
\affiliation{Department of Physics and Astronomy, University of Texas Rio Grande Valley, 
Brownsville, Texas 78520, USA}
\affiliation{Department of Physics, IIT Hyderabad, Kandai, Telangana-502284, India}

\author{Yong Shao}
\affiliation{School of Astronomy and Space Science, Nanjing University, Nanjing, 210023, China}
\affiliation{Key Laboratory of Modern Astronomy and Astrophysics, Nanjing University, Ministry of Education, Nanjing, 210023, China}

\date{\today}

\begin{abstract}

We evaluate the prospects for radio follow-up of the double neutron stars (DNSs) in the Galactic disk that could be detected through future space-borne gravitational wave (GW) detectors. 
We first simulate the DNS population in the Galactic disk that is accessible to space-borne GW detectors according to the merger rate from recent LIGO results. 
Using the inspiraling waveform for the eccentric binary, the average number of the DNSs detectable by TianQin (TQ), LISA, and TQ+LISA are 217, 368, and 429, respectively. For the joint GW detection of TQ+LISA, the forecasted parameter estimation accuracies, based on the Fisher information matrix, for the detectable sources can reach the levels of $\Delta P_{\mathrm b}/P_{\mathrm b} \lesssim 10^{-6}$, $\Delta \Omega \lesssim 100~{\mathrm {deg}}^2$, $\Delta e/e \lesssim 0.3$, and $\Delta \dot{P}_{\mathrm b} / \dot{P}_{\mathrm b} \lesssim 0.02$.
These estimation accuracies are fitted in the form of power-law function of signal-to-noise ratio. 
Next, we simulate the radio pulse emission from the possible pulsars in these DNSs according to pulsar beam geometry and the empirical distributions of spin period and luminosity. 
For the DNSs detectable by TQ+LISA, the average number of DNSs detectable by the follow-up pulsar searches using the Parkes, FAST, SKA1, and SKA are 8, 10, 43, and 87, respectively. 
Depending on the radio telescope, the average distances of these GW-detectable pulsar binaries vary from 1 to 7 kpc.
Considering the dominant radiometer noise and phase jitter noise, the timing accuracy of these GW-detectable pulsars can be as low as 70~${\rm ns}$ while the most probable value is about 100~$\mu {\rm s}$. 

\end{abstract}

\maketitle

%\end{CJK*} 

\section{Introduction}

Binary pulsars are emitters of both radio and gravitational waves (GWs) and, as has been the case since PSR B1913+16 \cite{HulseTaylor1975}, short-period binary pulsars provide excellent laboratories for constraining the theories of gravity \cite{2003LRR.....6....5S}. An excellent example is the Double Pulsar PSR J0737-3039A/B~\cite{Kramer2021}. Due to various technical limitations described below, current radio surveys for such systems can only find orbital periods longer than about an hour. However, upcoming space-borne GW detectors, i.e., LISA \cite{LISA2017,arXiv220306016}, TianQin (TQ) \cite{TianQin2016}, and Taiji \cite{10.1093/nsr/nwx116}, will be able to detect systems with shorter orbital periods. It is therefore interesting to consider the prospects of radio follow-up observations of short-period double neutron star (DNS) systems detected using GWs.

Through radio surveys, a total of 20 DNS systems have been discovered up to now \cite{Andrews2019,AMM2021}, in which the shortest orbital period is found in PSR J1946+2052 with a period of $1.88~{\rm hr}$ \cite{Stovall2018}. To detect pulsars in binary systems, pulsar searching algorithms, in principle, need to consider the Doppler smearing effect in arrival times of radio pulses due to orbital motion in addition to the routine de-dispersion, folding, radio frequency interference removal, etc. 
When integration time for pulsar searching, typically $10-30$~min, becomes a considerable fraction of the orbital period, the Doppler smearing  will be worsened and must be corrected in a searching algorithm. 
Several methods, such as acceleration search \cite{Faulkner2004} and jerk search \cite{Bagchi2013}, have been proposed to compensate for the loss of search sensitivity due to the binary motion, but they are only applicable for integration times that are shorter than tens of percent of the orbital period. The other methods, such as dynamic power spectrum \cite{Chandler2003} and phase modulation search \cite{Ransom2003}, have also been developed to extend the applicability to the case that the integration time can be approximately equal to or even longer than the orbital period, however, with a cost of the high demand of computational resources. 

As a consequence of the above limitations, all 359 binary pulsars found so far \cite{ATNF_website,Manchester2005} have orbital periods longer than $1.2~{\rm hr}$.
Tight Galactic DNS systems with orbital periods in the $3-60$~min range will emit GWs with frequencies in the mHz band accessible to space-borne GW detectors. 
From the discoveries by LIGO \cite{Abbott_2009} and Virgo \cite{Acernese_2008} of GWs from two DNS merger events, GW170817 \cite{GW170817} and GW190425 \cite{GW190425}, the inferred merger rate density is $10-1700 ~\mathrm{Gpc^{-3}~yr^{-1}}$ for a 90\% credible interval \cite{GWTC32021}. 
Taking $920 ~\mathrm{Gpc^{-3}~yr^{-1}}$ as the fiducial point estimate, as in \cite{GWTC12019}, we get a corresponding DNS merger rate of $210~\rm{Myr}^{-1}$ for a typical Milky-Way type galaxy. 
From this inferred merger rate, one expects $\sim 2100$ DNS systems with orbital periods range less than $1~{\rm hr}$  in our Galaxy \cite{Andrews2020}. 
Therefore, despite the challenges of detecting DNSs directly by radio surveys, the space-borne GW detectors can potentially detect DNS systems with orbital periods less than $1~{\rm hr}$. 

Several recent works have investigated the number of the DNSs in our Galaxy (and possible nearby galaxies) that can be detected by LISA. Seto estimated the frequency distribution of the DNSs in the Local Group of the galaxies based on the higher merger rate $1540~\mathrm{{Gpc}^{-3}{yr}^{-1}}$ derived from the combination of GW170817 and the LIGO-Virgo O1 data release \cite{GW170817}. This work shows that LISA can detect about 5 DNSs in the Large Magellanic Cloud and Andromeda with signal-to-noise ratios (SNRs) greater than 10 for a 10-year  observation \cite{Seto2019}. 
Lau et al. simulated a set of DNSs in the LISA band using the population synthesis code COMPAS \cite{COMPAS} and found that for a 4-year observation and the merger rate of $33~\mathrm{Myr}^{-1}$, LISA can detect about 35 DNSs with SNRs greater than 8 (33 in our Galaxy, 2 in the Large Magellanic Cloud and Small Magellanic Cloud) \cite{Lau2020}.
Andrews et al. used the merger rate of $210~\mathrm{Myr}^{-1}$ to obtain the number of the Galactic DNSs in the mHz band. They found that, for an observation of 4 (8) years and the SNR threshold of 7, LISA can detect 240 (330) DNSs in our Galaxy, 1 (4) in Andromeda \cite{Andrews2020}. 

The above studies have focused on the detection of DNSs using GWs alone. 
To mitigate the problems, outlined earlier, associated with detecting short-period DNSs using radio observations, Kyutoku et al. discussed a multi-messenger strategy to detect Galactic pulsar-NS binaries with orbital periods less than 10 min \cite{Kyutoku2019}. They showed that LISA can detect these DNSs with SNRs greater than 100 and accurately determine their sky positions and orbital parameters. 
In the scenario of synchronous observation of LISA and Square Kilometre Array (SKA) \cite{Dewdney2009} starting in the mid-2030s, the information provided by LISA will enable a reduction of 2 to 6 orders of magnitude in the number of telescopes pointing comparing with the all-sky search and a further reduction of 9 orders of magnitude in the number of trials for orbital modulation in order to correct the Doppler smearing.  
On the side of astrophysics, Thrane et al. explored the prospect of using the Lense-Thirring precession effect to constrain the equation of state of NSs based on LISA+SKA multi-messenger observations \cite{Thrane2020}. They showed that using 10-year observation of SKA the measurement accuracy of the mass-radius relation of NSs can reach $0.2\%$, which is an order of magnitude better than the X-ray pulsation measurements \cite{Riley_2019} and the GW observations \cite{Vivanco2019}.

Multi-messenger observations, including GW and radio (and possibly X-ray), can be highly complementary and provide a more complete picture of the DNS population and the structure of the NS. Regarding the DNS population, ground-based GW detectors have currently detected coalescing DNSs, while radio telescopes have detected DNSs with orbital periods of more than an hour. There is a gap in the detection of the DNSs with periods of a few minutes to an hour. The orbital periods of the DNSs that can be detected by space-borne GW detectors (e.g., LISA and TQ) can fill this gap. Once the pulsar binaries are detected from the GW-detectable DNSs, this will directly map the short-period pulsar binaries, which will help to study the DNS formation and population. Furthermore, with the detection of such a system we can study one of the most important problems of the NS, namely the equation of state. Although both GW detection and radio pulse detection can constrain the equation of state in separate ways and have their own detection ranges and accuracies, the information obtained from multi-messengers is expected to be able to constrain it more tightly \cite{Miller2020ApJ, Thrane2020}.

In this work, we study the detectability of Galactic disk DNSs in mHz frequency band using a network of space-borne GW detectors and the prospects for radio follow-up of such detected systems.
In addition to the number of detectable sources, this work forecasts the accuracies of the estimated parameters of the DNSs by taking into account the antenna responses of space-borne GW detectors to the inspiraling waveforms of these systems. 
The follow-up search for the possible radio pulsar components in the DNS systems is studied based on both the intrinsic properties of the pulsars and the specifications of the large radio telescopes including the Parkes radio telescope \cite{2020PASA...37...12H}, the Five-hundred-meter Aperture Spherical radio Telescope (FAST) \cite{2011IJMPD..20..989N}, SKA phase 1 (SKA1) \cite{dewdney2016} and SKA \cite{2020PASA...37....2W}. 
We estimate the timing precision of the detectable pulsars by the radio telescopes, which are crucial for the further investigations of the ultra-compact DNSs as laboratories for astrophysics and fundamental physics. 

The rest of this paper is organized as follows. Sec.~\ref{sec:DNSpopulation} presents the simulation of the DNS population in the Galactic disk.  Sec.~\ref{sec:waveform_noise_PEA} introduces the GW signal model for the DNS in an eccentric orbit, the noise characteristics for LISA and TQ, and the Fisher information approach to obtain the parameter estimation accuracy. The numbers of detectable DNSs and parameter estimation accuracies expected for TQ, LISA, and TQ+LISA network are discussed in  Sec.~\ref{sec:GWresults}. 
Sec.~\ref{sec:surveysimulation} investigates the follow-up searches of the pulsars based on the Parkes, FAST, SKA1, and SKA. The pulsar beam geometry and the empirical distributions of the spin periods and luminosities are given in Sec.~\ref{beammodel_distribution}. The synergy for the multi-messenger observations with the GW detectors and the radio telescopes is given in Sec.~\ref{pulsarresult}. The paper concludes in Sec.~\ref{sec:conclusion}. 

\section{Galactic DNS population} \label{sec:DNSpopulation}

In this work, we follow the method in Sec. 2.4 of \cite{Andrews2020} to synthesize the Galactic population of tight DNSs. 
We assume the component masses of all DNSs to be $m_1 = m_2 = 1.4~M_\odot$. Hence, the chirp masses of the binaries are $M_{\mathrm c} = (m_1 m_2)^{3/5}/(m_1+m_2)^{ 1/5} = 1.22 ~M_\odot$.

Given the DNS merger rate and the Galactic disk model, we obtain the number and locations of the DNSs at a specific evolution time. Here, we focus on those DNSs that will merge within $10~{\rm Myr}$. 
%(the corresponding orbital period $P_{\rm b} \lesssim 1~{\rm hr}$). 
Note that there is no need to extend to longer merger times because the sensitivity of LISA or TQ is limited to DNSs that are expected to merge within the next $\sim 10~\mathrm{Myr}$ (cf. Fig.~\ref{fig:DNSchara}).
Using the merger rate $R_{\mathrm{MW}} = 210 ~\mathrm{Myr}^{-1}$ \cite{Andrews2020,GWTC12019}, the expected number of DNSs in our simulation is $N_\mathrm{DNS}= R_{\mathrm{MW}} \times 10 ~\mathrm{Myr} = 2100$.
We assume that these DNSs are randomly distributed in the Galactic disk with a number density profile  \cite{Nelemans2001}
\begin{equation}\label{Gdiskdist}
f_{\mathrm{dis}} (r, z) \propto e^{-r / L} \mathrm{sech\,}(z / \beta)^{2} ~\mathrm{pc}^{-3} \,,
\end{equation}
where the disk scale length $L=2.5~\mathrm{kpc}$, the scale height $\beta=0.2~\mathrm{kpc}$. The distribution for the azimuthal angle follows $\mathcal{U}[0, 2\pi]$. Here, $\mathcal{U}(a,b)$ denotes the continuous uniform distribution within the interval $[a,b]$. 

The differential equations of the binary orbit due to GW emission allow us to obtain the evolution of these parameters with time.
For the binary in an eccentric orbit, the time evolution of its orbital semi-major axis $a$ and eccentricity $e$ due to GW emission can be expressed as \cite{Peters1964}:
\begin{eqnarray}
\left\langle\frac{{\rm d} a}{{\rm d}t}\right\rangle &=& -\frac{64}{5} \frac{G^{3} \mu M^2}{c^{5} a^{ 3}} F(e) \label{eq:dadt} \,, \\ 
\left\langle\frac{{\rm d}e}{{\rm d}t}\right\rangle &=& -\frac{304}{15} \frac{G^{3} \mu M^2 e}{c^{5} a^ {4}\left(1-e^{2}\right)^{5 / 2}}\left(1+\frac{121}{304} e^{2}\right) \,, \label{eq:dedt}
\end{eqnarray}
where $\left\langle \cdot \right\rangle$ denotes the average over one orbital period. $M$ and $\mu$ are the total mass and reduced mass of the system. 
\begin{equation}
F(e)={\left( 1+\frac{73}{24}{{e}^{2}}+\frac{37}{96}{{e}^{4}} \right)} /{{{\left( 1-{{e}^{2}} \right)}^{7/2}}}
\end{equation}
is the enhancement factor. Combining Eq. (\ref{eq:dadt}) and Eq. (\ref{eq:dedt}), one gets the evolution of the semi-major as a function of the orbit eccentricity: 
\begin{equation} \label{eq:a(e)}
a(e) = c_{0} \frac{e^{12 / 19}}{1-e^{2}}\left(1+\frac{121}{304} e^{2}\right) ^{870/2299} \,.
\end{equation}
Let $g(e) = a(e)/c_0$, then $c_0$ satisfies $a(e_0) = c_0 g(e_0)$. 

One can solve the evolution of $e$ over time by inserting Eq.~(\ref{eq:a(e)}) in Eq.~(\ref{eq:dedt}) and numerically integrating the ordinary differential equation. Since Eq.~(\ref{eq:a(e)}) contains $c_0$, the evolution of $e$ over time depends on the initial (or reference) values $a_0$ and $e_0$, as well as the integration time.
% (evolution time or merger time) from the initial state to the merger.
We take the orbital parameters of PSR B1913+16 close to the merger as the reference values: $P_\mathrm{b0} = 0.2 ~\mathrm{s}$, $e_0 = 5 \times 10^{-6}$ \cite{Andrews2020}. 
We draw the merger time of the DNSs in our simulations from a uniform distribution over the past $10~\mathrm{Myr}$.
Combining Kepler's third law with Eq.~(\ref{eq:a(e)}), one can obtain the relation between $e$ and orbital period $P_\mathrm{b}$, then the evolution of $P_\mathrm{b}$ over time can also be obtained. 

The DNS population is generated by assigning these orbital parameters to the aforementioned simulated sources.

\section{Detecting GW of DNS}
\label{sec:GWdetection}

\subsection{DNS gravitational waveform, detector noise, and parameter estimation accuracy}
\label{sec:waveform_noise_PEA}

Under the transverse-traceless gauge and quadrupole moment approximation, we can analytically deduce the gravitational waveforms for two polarizations of an eccentric binary \cite{Moreno-Garrido1995}
\begin{widetext}
\begin{eqnarray}\label{eq:hplushcross}
h_{+} &=& -\frac{h_0}{2} \sum_{n=1}^{\infty} \left\{ \sin^{2} \iota  J_{n}(n e)  \cos n M_{\mathrm{A}} + \left(1+\cos^{2} \iota\right) 
 \left[\frac{C_{n}-S_{n}}{2} \cos \left(n M_{\mathrm{A}}+2 \Phi_{\mathrm{P}}\right) + \frac{S_{n}+C_{n}}{2} \cos \left(n M_{\mathrm{A}}-2 \Phi_{\mathrm{P}}\right)\right] \right\}  \,, \nonumber \\ 
h_{\times} &=& -h_0 \cos \iota \sum_{n=1}^{\infty} \left[ {\frac{S_{n}-C_{n}}{2}   \sin \left(n M_{\mathrm{A}}+2 \Phi_{\mathrm{P}}\right)} +  \frac{S_{n}+C_{n}}{2} \sin \left(n M_{\mathrm{A}}-2 \Phi_{\mathrm{P}}\right)  \right]  \,, 
\end{eqnarray}
\end{widetext}
with  
\begin{eqnarray}
h_0 &=&  \frac{4 \left(G M_{\mathrm{c}}\right)^{5/3} \omega_{\mathrm{b}}^{2 / 3}}{c^4 D} \label{eq:amp}  \,, \\
{{S}_{n}} &=& -\frac{2{{\left( 1-{{e}^{2}} \right)}^{{1}/{2}}}}{ne}\frac{{\rm d}{{J}_{n}}(ne)}{{\rm d}e}+\frac{2n{{\left( 1-{{e}^{2}} \right)}^{{3}/{2}}}{{J}_{n}}(ne)}{{{e}^{2}}} \label{eq:sn} \,, \\
{{C}_{n}} &=& -\frac{\left( 2-{{e}^{2}} \right){{J}_{n}}(ne)}{{{e}^{2}}}+\frac{2\left( 1-{{e}^{2}} \right)}{e}\frac{{\rm d}{{J}_{n}}(ne)}{{\rm d}e}  \,. \label{eq:cn}
\end{eqnarray}
Here ${{\omega }_{\mathrm{b}}}={2\pi }/{{{P}_{\mathrm{b}}}}$ is the orbital angular frequency, $D$ is the distance to the source, $\iota$ is the inclination angle of the orbit, ${{J}_{n}}(ne)$ is the Bessel function of the first kind $(n=1, 2, 3, ...)$ with $n$ as the harmonic number, ${{M}_{\mathrm{A}}}={2\pi (t-{{t}_{0}})}/P_{\mathrm{b}}$ is the mean anomaly, $ {{t}_{0}}$ is the time of periastron passage, and ${{\Phi }_{\mathrm{P}}}$ is the periastron phase.
%By introducing the frequency derivative and Doppler phase modulation in the GW phase, the above signal can be transformed as follows to include the corresponding effects 
The aforementioned waveforms can be transformed as follows to include the effects of the orbital frequency derivative and Doppler phase modulation in the GW phase: 
\begin{eqnarray}
n M_{\mathrm{A}} \longrightarrow ~&& n M_{\mathrm{A}} - n \frac{\pi \dot{P_\mathrm{b}}}{P_\mathrm{b}^2} (t-t_0)^2 + \nonumber \\
&& 2 \pi \frac{n}{P_\mathrm{b}} \frac{R}{c} \sin \theta_{\mathrm{S}} \cos \left(2 \pi f_{m} t-\phi_{\mathrm{S}} \right)  \,,
\end{eqnarray}
where $R$ is the distance between the Sun and Earth and $f_m=1/(1~{\rm yr})$ is the  modulation frequency due to the annual revolution of the detector's constellation (for either LISA or TQ) around the Sun. ($\theta_{\mathrm{S}},\phi_{\mathrm{S}}$) are the colatitude and longitude of the source in the heliocentric ecliptic coordinate system. 

At the low frequency (i.e., long wavelength) limit, the strain signal induced by GWs can be expressed as 
\begin{equation}\label{eq:strain}
h_\mathrm{I, II}(t)={{F}^{+}_\mathrm{I, II}}(t)\cdot {{h}_{+}}(t)+{{F}^{\times }_\mathrm{I, II}}(t)\cdot {{h}_{\times }}(t)  \,,
\end{equation}
where ${{F}^{+,\times }_\mathrm{I, II}}$ is the antenna response functions of the detector for the two polarizations ($+,\times$) and two independent Michelson-type  interferometers (I, II) in the heliocentric ecliptic coordinates. %  $(D,\theta_{\mathrm{S}},\phi_{\mathrm{S}})$,
See \cite{Cornish2003} and \cite{Hu2018} for their explicit expressions for LISA and TQ, respectively.

The non-sky-averaged sensitivity curves for LISA \cite{Robson2019} and TQ \cite{Huang2020} can be obtained by 
%
% \begin{widetext}
\begin{eqnarray}\label{eq:LISAsen}
{{S}_{\rm L}}(f) &=& \frac{1}{{{L}_{\rm L}^2}}\left[ {{P}_{\mathrm{OMS}}}(f)+\frac{4{{P}_{\mathrm{acc}}}(f)}{{{\left( 2\pi f \right)}^{4}}} \right]\left[ 1+\frac{6}{10}{{\left( \frac{f}{{{f}_{*}^{\rm L}}} \right)}^{2}} \right] 
\nonumber \\
&& + {{S}_{c}}(f) \,, \\
{{S}_{\rm T}}(f) &=& \frac{1}{{{L}_{\rm T}^2}}\left[ \frac{4{{S}_{a}}}{{{\left( 2\pi f \right)}^{4}}}\left( 1+\frac{{{10}^{-4}}~\mathrm{ Hz}}{f} \right)+{{S}_{x}} \right] \nonumber \\
&& \times \left[ 1+0.6{{\left( \frac{f}{{{f}_{*}^{\rm T}}} \right)}^{2}} \right]  \,.
\end{eqnarray}
% \end{widetext}
%
The arm length for LISA and TQ are 
${L}_{\rm L} = 2.5 \times 10^9~{\rm m}$ and ${L}_{\rm T} = 1.7 \times 10^8~{\rm m}$, respectively. 
The associated transfer frequency for them are 
${f}_{*}^{\rm L} = 19.09~{\rm {mHz}}$ and ${f}_{*}^{\rm T} = 0.28~{\rm {Hz}}$.
The detailed budgets for single-link optical metrology noise ${P}_{\mathrm{OMS}}$ and single test mass acceleration noise ${P}_{\mathrm{acc}}$ for LISA and corresponding noise ${S}_{x}$ and ${S}_{a}$ for TQ can be found in 
\cite{Robson2019} and \cite{Hu2018,Huang2020}, respectively. 
$S_c$ is the PSD of the stochastic GW signal associated with the unresolvable Galactic binary population as a whole (dominated by WD binaries).
Note that we do not include $S_{c}$ for TQ since the instrumental noise for TQ is higher than it in the concerned frequency range. 

Assuming the noise of LISA and TQ are uncorrelated, the joint sensitivity curve of LISA and TQ can be obtained from the following expression \cite{Tinto2016}
\begin{equation}\label{eq:sntl}
{S_{\rm {T + L}}}(f) \equiv \frac{{S_{\rm T}}(f){S_{\rm L}}(f)}{{S_{\rm T}}(f) + {S_{\rm L}}(f)} \,.
\end{equation}

Given GW signal is sufficiently strong, the estimation accuracies of unknown parameters in the GW signal
%, i.e., these included in Eq.~\ref{eq:strain}, 
can be approximately calculated by using the Fisher information matrix (FIM),
\begin{equation}
{{\Gamma }^{ij}} \equiv \left( \frac{\partial h}{\partial {{\lambda }_{i}}},\frac{\partial h}{\partial {{\lambda }_{j}}} \right)  \,, 
\end{equation}
where (,) denotes the noise-weighted inner product that contains both the GW signal and the detector noise \cite{Finn1992}. 
Similarly, the optimal SNR for signal detection is defined as ${\rm SNR}\equiv (h,h)^{1/2}$.
The root-mean-square (rms) error of parameter ${\lambda_i}$ is estimated as
\begin{equation}
\Delta {\lambda_i} = \sqrt {{\Sigma_{ii}}} \,,
\end{equation}
where the variance-covariance matrix $\Sigma$ is the inverse of the FIM, i.e., $\Sigma  = {\Gamma^{-1}}$.
The correlation coefficient
between $\lambda_i$ and $\lambda_j$ can be calculated by 
\begin{equation}\label{eq:cc}
c_{ij} = \Sigma_{ij} / \sqrt{\Sigma_{ii}\Sigma_{jj}} \,.
\end{equation}

For a GW source with sky location $(\theta_\mathrm{S},\phi_\mathrm{S})$, its sky localization error is defined as \cite{cutler1998lisa}
\begin{equation}\label{eq:DeltaOmega}
\Delta \Omega = 2 \pi \sqrt { \Sigma_{\cos \theta_\mathrm{S} \cos \theta_\mathrm{S}} \Sigma_{\phi_\mathrm{S} \phi_\mathrm{S}}-\Sigma_{\phi_\mathrm{S} \cos \theta_\mathrm{S}}^{2} }  \,.
\end{equation}

\subsection{Results on detectable DNSs}
\label{sec:GWresults}

We assume the observation time $T_{\mathrm {obs}} = 4 ~\mathrm{yr}$. 
The inclination angle $\cos\iota \sim \mathcal{U}(- 1,1)$, the polarization angle ${\psi} \sim \mathcal{U}(0,\pi)$, and the periastron phase $\Phi_{\mathrm P} \sim \mathcal{U}(0,2\pi)$. 
%Here, $\mathcal{U}(a,b)$ denotes the continuous uniform distribution within the interval $[a,b]$.
We draw $N_\mathrm{DNS}$ random samples from the above distributions and assign them to the simulated DNSs in the Galactic disk.
The assumed maximum unknown parameter space for FIM is $\bm {\lambda} = \left(\ln h_0, \cos \iota, \psi, \cos \theta_\mathrm{S}, \phi_\mathrm{S}, \ln P_\mathrm {b}, \Phi_\mathrm{P}, \ln e, \ln \dot {P}_\mathrm{b} \right)$. We set the SNR threshold to be 7, above which we regard the source as detectable.  According to the chirp frequency limit \cite{Nelemans2001}, i.e., $\dot f T_{\mathrm{obs}} \geqslant 1/T_{\mathrm{obs}}$, we get the frequency range in which the chirp mass can be measured through GW signals 
%
%\begin{small}
\begin{equation}\label{eq:fchirp}
\frac{f_{\mathrm {chirp }}}{\mathrm{~mHz}} \geqslant 1.75 \left(\frac{{M}_\mathrm{c}}{1.22 ~M_{\odot}}\right)^{-5 / 11}\left(\frac{T_{\mathrm{obs}}}{4 ~\mathrm{yr}}\right)^{-6 / 11} F(e)^{-3/11} \,.
\end{equation}
%\end{small}
Note that chirp mass and eccentricity are degenerate in the GWs of DNS when the eccentricity is non-negligible. In this case, Eq.~(\ref{eq:fchirp}) is a necessary but not sufficient condition for the measurement of chirp mass. However, as will become clear later, in the current work, Eq.~(\ref{eq:fchirp}) only serves as a criterion for selecting sources for subsequent evaluation of the estimation accuracy of chirp mass. 

For the DNSs with SNRs higher than the threshold but the GW frequency lower than $f_{\mathrm {chirp }}$, their parameter space is $\bm {\lambda}$ with $ \ln \dot {P }_\mathrm{b}$ eliminated. 

For a quasi-monochromatic source with frequency $f_0$ and observation time $T_\mathrm{obs}$, the FIM can be simplified as \cite{Cutler1998,Shah2012}
\begin{eqnarray} %\label{eq:gammasnr}
{\Gamma }^{ij} &=& \frac{2}{S_{n}\left(f_{0}\right)} \sum_{\alpha=\mathrm{I, II}} \int_{-\infty}^{\infty} \partial_{i} \widetilde{h}_{\alpha}^{*}(f) \partial_{j} \widetilde{h}_{\alpha}(f) df  \,,  \nonumber \\
&\simeq& \frac{2}{S_{n}\left(f_{0}\right)} \sum_{\alpha=\mathrm{I, II}} \int_{0}^{T_\mathrm{obs}} \partial_{i} h_{\alpha}(t) \partial_{j} h_{\alpha}(t) dt \,. \label{eq:gammaapprox}
\end{eqnarray}
Here $\widetilde{h}_{\alpha}^{*}(f)$ is the complex conjugate of the Fourier transform of $h_{\alpha}(t)$. 
Note that $\mathrm{SNR}^2 = \Gamma ^{11}$ for the parameter space we choose. We can replace the partial derivatives in FIM with the central finite difference  
\begin{equation}
\frac{\partial h}{\partial \lambda_{i}} \approx \frac{h\left(t, \lambda_{i}+\delta \lambda_{i}\right)-h\left(t, \lambda_{i}-\delta \lambda_{i}\right)}{2 ~\delta \lambda_{i}} \,,
\end{equation}
where $\delta \lambda_i/\lambda_i \lesssim 10^{-8}$, and their values can be adjusted to obtain stable results \cite{Shah2012}. In addition, the time integral in Eq.~(\ref{eq:gammaapprox}) is transformed into a summation in which the sampling frequency is selected to be $f_{\mathrm s} = 4f_0$.

\begin{figure}[!htbp]
  \centering
  \includegraphics[scale=0.55]{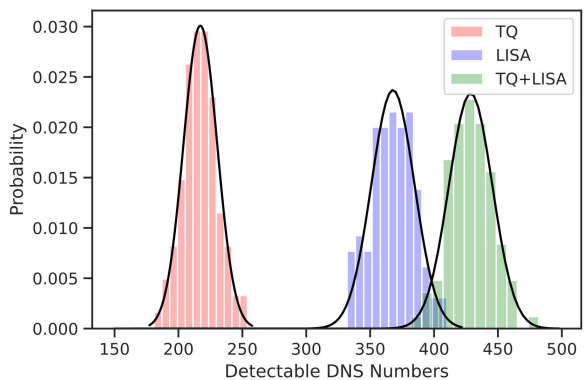}
  \caption{The histograms for the number of detectable DNSs by TQ, LISA, and TQ+LISA network from 100 independent Monte Carlo simulations. Black curves are the best-fit Gaussian distributions.} 
  \label{fig:GW_Detectable_DNS_numbers}
\end{figure}

We perform 100 independent Monte Carlo simulations of the DNSs in the Galactic disk and use Gaussian distributions to fit the numbers of the DNSs detectable by TQ, LISA, and TQ+LISA network (ideally this indicates that TQ and LISA will be operated synchronously and the data are analyzed jointly), respectively (Fig.~\ref{fig:GW_Detectable_DNS_numbers}). The average numbers of detectable DNSs are 217, 368, and 429 for those three cases. 

\subsubsection{Characteristic strain}

\begin{figure}[htbp]
\centering
\includegraphics[scale=0.46]{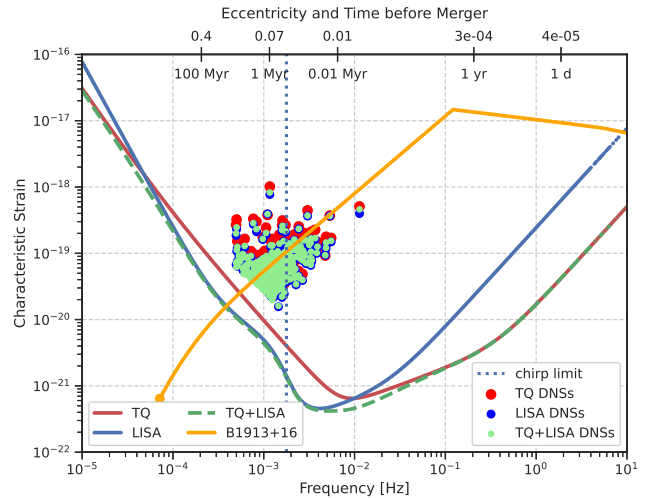}
\caption{The characteristic strain of the simulated DNSs from one of the Monte Carlo simulations. The bottom abscissa is the GW frequency of the 2nd harmonics $f_{n=2}$, and the top abscissa is the evolution of the orbital eccentricity of the DNSs and the corresponding time before its merger based on the parameters of the benchmark source PSR B1913+16. 
The red solid, blue solid and green dashed curves represent the characteristic sensitivity of TQ, LISA, and TQ+LISA, respectively. The red, blue, and green dots represent the detectable DNSs by TQ, LISA, and TQ+LISA. The blue vertical dotted line is the chirp frequency limit for a circular orbit, above which the evolution of the frequency is detectable.}
\label{fig:DNSchara}
\end{figure}

In Fig.~\ref{fig:DNSchara}, we plot the characteristic strains for signal ($h_{\mathrm{c},n}$) and noise ($h_{{\mathrm{noise},n}}$) which are defined by rewriting the expression of SNR as follows,
\begin{eqnarray}\label{eq:snrn}
\mathrm{SNR}^2 &=& \sum_{n=1}^{\infty} \mathrm{SNR}^2_n = \sum_{n=1}^{\infty} \frac{h^2_{\mathrm{c},n} }{h^2_{{\mathrm{noise},n}}} \,, \nonumber \\
&=& \sum_{n=1}^{\infty}  \frac{f_n \int_{0}^{T_\mathrm{obs}} 2\left(h_{\mathrm{I},n}^{2}+h_{\mathrm{II},n}^{2}\right) dt}{f_n S(f_n)} \,, 
\end{eqnarray}
where the subscript $n$ represents the $n$-th harmonics of the GW in Eq.~(\ref{eq:hplushcross}). Each source includes the influence of the antenna response functions.
Similarly, the joint characteristic strain of TQ and LISA can also be derived from the combined SNR with the joint sensitivity in Eq.~(\ref{eq:sntl}). 

Given the orbital parameters of the reference source PSR B1913+16 and merger time distribution $\mathcal{U}[0, 10]~\mathrm{Myr}$), the maximum eccentricity ($\approx 0.18$) in our simulated DNSs is relatively low. Thus, the GW signal from the 2nd harmonic is dominant \cite{Moreno-Garrido1995}, so we only draw the characteristic strain from the 2nd harmonic in Fig.~\ref{fig:DNSchara}. However, in our calculation of the SNR and FIM, the first four harmonics ($n=1,2,3,4$) are taken into account, while the cross terms between different harmonics are neglected \cite{Mikoczi2012}.

Since the benchmark used in the calculation of the orbital parameters of the DNS comes from binary pulsar PSR B1913+16 at merger ($P_\mathrm{b0} = 0.2 ~\mathrm{s},\, e_0 = 5 \times 10^{-6}$), we add the characteristic strain from the 2nd harmonic of this binary in Fig.~\ref{fig:DNSchara} as a reference \cite{Andrews2020}.

\subsubsection{Parameter estimation accuracy}

\begin{figure}[!htbp]
\centering
\includegraphics[scale=0.42]{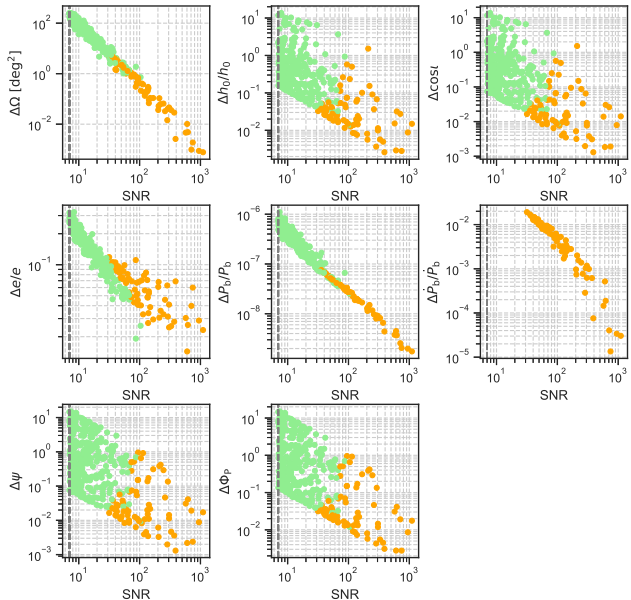}
\caption{The rms errors of the parameters for the DNSs from one of the Monte-Carlo simulations. All sources have an SNR of the TQ+LISA network larger than the threshold (vertical dashed lines near the left edges). 
% In each panel, the green dots represent the sources that have the 8D parameter space, while the orange dots have the 9D parameter space (the sky location combines $\theta_\mathrm{S}$ and $\phi_\mathrm{S}$ as in Eq.~(\ref{eq:DeltaOmega})).
The green dots correspond to relatively low-frequency systems with $\dot{P}_{\mathrm b}$ set to zero and excluded from the parameter estimation space.
Overall, the parameter estimation accuracies improve as the SNR increases. There is a strong anti-correlation between the amplitude and the inclination as well as the polarization angle and the periastron phase, hence their distributions show larger scattering. }
\label{fig:errorsnr8}
\end{figure}

As shown in Fig.~\ref{fig:errorsnr8}, the rms errors of all parameters in the FIM decrease, in general, with the increase of SNR. 
Among them, the relative estimation accuracy of the orbital period (or frequency) is the highest, $\Delta P_{\mathrm b}/P_{\mathrm b} \lesssim 10^{-6}$. 
% that means the determination of frequency is the most accurate. 
It is clear that there are strong correlations between the amplitude and the inclination angle as well as the polarization angle and the periastron phase. %, and their estimation accuracy is almost the same. 
For a source with SNR around 100, the sky localization $\Delta \Omega \lesssim 1~{\mathrm {deg}}^2$, the relative error of orbital eccentricity $\Delta e/e \lesssim 0.1$, the relative error of the orbital period derivative $\Delta \dot{P}_{\mathrm b} / \dot{P}_{\mathrm b} \lesssim 0.004$. As represented by the orange dots, the orbital period derivative $\dot{P}_{\mathrm b}$ can be measured in about $17\%$ of all sources. As shown below, the measurement of $\dot{P}_{\mathrm b}$ is critical for the subsequent determination of the chirp mass and source distance. 

Following the analysis of the correlation between parameters in \cite{Shah2012}, the strain waveforms in Eq.~(\ref{eq:strain}) with inclination $0^{\circ} \le \iota \lesssim 45^{\circ}$ have the same waveform structure, so by changing the amplitude or the inclination, we can obtain a very similar signal, i.e. there is a strong anti-correlation between the $h_0$ and $\cos{\iota}$. However, for inclination $45^{\circ} \lesssim \iota \le 90^{\circ}$, the magnitude and structure of the signal change for different inclinations, so they are no longer correlated. Secondly, the polarization angle $\psi$ is measured in the plane perpendicular to the GW propagation direction, while the periastron phase angle $\Phi_{\rm p}$ is measured in the orbital plane. When the sources are getting close to being face-on to the detector, $\psi$ and $\Phi_{\rm p}$ will appear to have correlations due to partial overlap between those two planes. This explains the poor parameter accuracy for some detectable sources (see Fig.~\ref{fig:errorsnr8}).

\begin{figure}[!htbp]
\centering
\includegraphics[scale=0.42]{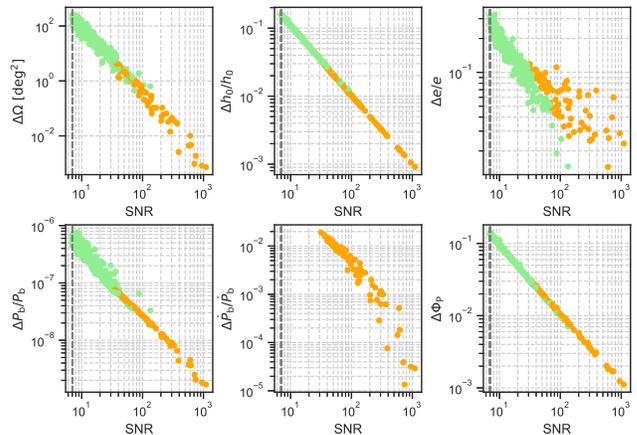}
\caption{Same as Fig.~\ref{fig:errorsnr8}, but remove the inclination $\iota$ and polarization angle $\psi$ in the two pairs of strongly correlated parameters. }
\label{fig:errorsnr6}
\end{figure}

In the case that the electromagnetic observations can provide an accurate measurement of the inclination angle (so we remove the inclination angle $\iota$ from the FIM), the amplitude $h_0$ can then be well constrained through GW observation (see Fig.~\ref{fig:errorsnr6}). Likewise, this is also the case for the periastron phase and polarization angle. By comparing corresponding panels in Fig.~\ref{fig:errorsnr8} and  Fig.~\ref{fig:errorsnr6}, we can see that, when one of a pair of strongly correlated parameters is removed from the FIM, the error of the other parameter will reduce to the lower boundary of its widely dispersed distribution in Fig.~\ref{fig:errorsnr8}. 

The fitting result for the parameter estimation accuracy that does not include the frequency derivative (see the green dot in Fig.~\ref{fig:errorsnr6}) is 
\begin{eqnarray}
\Delta \Omega &\sim& 90 \left(\frac{10}{\mathrm{SNR}}\right)^{2} \mathrm{deg^2}  \,, \\
\frac{\Delta h_0}{ h_0} &\sim& 0.1 \left(\frac{10}{\mathrm{SNR}}\right) \,, \\
\frac{\Delta e }{ e} &\sim& 0.2 \left(\frac{10}{\mathrm{SNR}}\right) \,, \\
\frac{\Delta P_{\mathrm b} }{ P_{\mathrm b}} &\sim& 3.8 \times 10^{-7}\left(\frac{10}{\mathrm{SNR}}\right) \,, \\
\Delta \Phi_{\mathrm P} &\sim& 0.1 \left(\frac{10}{\mathrm{SNR}}\right) \mathrm{rad} \,.
\end{eqnarray}

When the orbital period (or frequency) derivative is included in the FIM, except for the eccentricity, the other parameters still maintain a power law of the SNR and hold the same as before. For $30 \lesssim \mathrm{SNR} \lesssim 100$, the relative errors of eccentricity and the orbital period derivative are roughly inversely proportional to the SNR. But when $\mathrm{SNR} \gtrsim 100$, their results tend to have larger dispersion. The significant dispersion of the relative error of $e$ is caused by the interplay between the detector's sensitivities at different frequencies and the evolution of $e$, which results in a larger dispersion of $e$ and subsequently a larger dispersion of $\Delta e/e$ for larger SNRs than for lower SNRs.
\begin{equation} \label{Deltae2e}
\frac{\Delta e}{e} \sim\left\{\begin{array}{ll}
0.08  \left(\frac{50}{\mathrm{SNR}}\right), \quad \mathrm{SNR} \lesssim 100 \\
0.04  \left(\frac{200}{\mathrm{SNR}}\right)^{0.3},  \quad \mathrm{SNR} >100
\end{array}\right.
\end{equation}
\begin{equation}
\frac{\Delta \dot{P}_{\mathrm b}}{\dot{P}_{\mathrm b}} \sim \left\{\begin{array}{l}
0.01 \left(\frac{50}{\mathrm{SNR}}\right), \quad \mathrm{SNR} \lesssim 100 \\
0.001 \left(\frac{200}{\mathrm{SNR}}\right)^{2}, \quad \mathrm{SNR} >100
\end{array}\right.
\end{equation}

Using the standard deviation propagation similar to that in \cite{Berti2005}, for the chirping sources, the relative error of the chirp mass and the distance of the source can be derived by
\begin{small}
\begin{equation}\label{deltamc}
\frac{\Delta M_{\rm c}}{M_{\rm c}} = \left[ \left(\frac{11}{5} \frac{\Delta P_{\rm b}}{P_{\rm b}} \right)^{2} + \left(\frac{3}{5} \frac{\Delta \dot P_{\rm b}}{\dot P_{\rm b}}\right)^{2}
+ \left(\frac{3}{5} \frac{d F(e)}{d e} \frac{\Delta e}{F(e)} \right)^{2} \right]^{1/2}  \,,
\end{equation}
\end{small}
and 
\begin{small}
\begin{equation}
\frac{\Delta D}{D} = \left[ \left(\frac{5}{3} \frac{\Delta M_{\rm c}}{M_{\rm c}}\right)^{2}+\left(\frac{2}{3} \frac{\Delta P_{\rm b}}{P_{\rm b}}\right)^{2}+\left(\frac{\Delta h_{0}}{h_{0}} \right)^{2} \right]^{1/2}  \,.
\end{equation}
\end{small}

The fitting equations for the above parameters are
\begin{eqnarray}
\frac{\Delta M_{\rm c}}{M_{\rm c}}
% &\approx& \frac{3}{5} \frac{\Delta \dot P_{\rm b}}{\dot P_{\rm b}} 
&\sim & \left\{\begin{array}{l}
0.007 \left(\frac{50}{\mathrm{SNR}}\right), \quad \mathrm{SNR} \lesssim 100 \\
0.001 \left(\frac{200}{\mathrm{SNR}}\right)^{2}, \quad \mathrm{SNR} >100
\end{array}\right. \\
\frac{\Delta D}{D} &\approx& \frac{\Delta h_{0}}{h_{0}} \sim 0.1 \left(\frac{10}{\mathrm{SNR}}\right)  \,.
\end{eqnarray}
%
% The observed mass distribution of the Galactic DNSs can place strong constraints on their evolution model \cite{Vigna-Gomez2018}.

With the small relative error in the estimation of chirp mass ($\Delta M_{\rm c} /M_{\rm c} \lesssim 1\%$), one may distinguish DNS ($M_{\rm c}=1.22~M_\odot$) from typical WD-NS binary ($M_{\rm c}=0.78~M_\odot$). But for a few extremely WDs with $M_{\rm c} \lesssim 1.2~M_{\odot}$, as discussed by \cite{Lau2020}, these WD binaries still can not be distinguished from DNSs. A general constraint can be made from the binary evolution models, such as the eccentric orbit and the disc binaries favor the DNSs. The electromagnetic follow-up around the estimated sky localization can provide further evidence to discern these binaries. As an example, we discuss the follow-up by radio telescopes in the next section.

\section{Radio follow-up of pulsars}
%\section{Simulations of pulsar surveys}
\label{sec:surveysimulation}

In this section, we evaluate the numbers and distances of GW-detectable DNSs in Sec.~\ref{sec:GWdetection} that can be detected as radio pulsar binaries based on the pulsar beam geometry and the detection conditions. The measurement accuracies of the time of arrivals (TOAs) for the detectable pulsars are estimated by considering the dominant sources of noise, namely, radiometer noise and phase jitter noise. 

\subsection{Pulsar beam geometry and empirical distributions} 
\label{beammodel_distribution}

The detectability of a pulsar depends on the orientation of the pulsar's spin axis, magnetic axis, and the observer's line of sight. 
In the pulsar beam geometry, the intrinsic pulse width $W$ measured in longitude of rotation satisfies the following relationship \cite{Gil1981, Gil1984} 
\begin{equation}\label{eq:pulsewidth}
\sin ^{2}\left(\frac{W}{4}\right) = \frac{\sin ^{2}(\rho / 2)-\sin ^{2}((\xi - \alpha) / 2)}{\sin \alpha \sin \xi} \,,
\end{equation}
where $\alpha$ is the angle between the spin axis and the magnetic axis, $\xi$ is the angle between the spin axis and the line of sight, and $\rho$ is the angular radius of the cone-shaped beam that is centered on the magnetic axis.  
Note that $\xi-\alpha$ is usually defined as the impact angle which is the closest approach between the magnetic axis and the line of sight when the former is rotating around the spin axis \cite{2004hpa..book.....L}. 

According to \cite{Gil1996}, the angle $\xi$ follows a probability distribution, $\cos\xi \sim \mathcal{U}(0,1)$, and $\alpha \sim \mathcal{U}(0,\pi/2)$. 
For a cone in a circular shape, the condition under which the radio beam can sweep across the line of sight is 
\begin{equation}\label{eq:beta_rho}
|\xi - \alpha| < \rho  \,.
\end{equation}
Otherwise, the beam will miss the observer. In the case that Eq.~(\ref{eq:beta_rho}) is satisfied while the right-hand side of Eq.~(\ref{eq:pulsewidth}) is in the range of $[0,1]$, the intrinsic pulse width $W$ can be obtained from Eq.~(\ref{eq:pulsewidth}). 

It has been found that the distribution of $\rho$ is strongly correlated with pulsar's spin period $P$, which follows an empirical formula \cite{Kramer1998}
\begin{equation}\label{rhoperiod}
\rho=\left\{\begin{array}{l}
5.4^{\circ} (P/{\rm{s}})^{-1 / 2},\quad P>30 \mathrm{~ms} \,, \\
31.2^{\circ},\quad P \leq 30 \mathrm{~ms} \,.
\end{array}\right.  
\end{equation}
To account for the fluctuations in the actual situation, in addition to the fiducial value $\bar{\rho}$ obtained from the above equation, the observed value of $\rho$ needs to add a random component $p$,   
\begin{equation}\label{rhoperiod+p}
\log \rho=\log \bar{\rho} +p  \,.
\end{equation}
Here, $p\sim \mathcal{U}(-0.15,0.15)$\cite{Smits2009}. 

By fitting the pulsars found in the Parkes Multibeam Pulsar Survey (PMPS) and the Parkes High-latitude Pulsar Survey (PHPS), the probability distribution of the spin periods for the isolated normal pulsars (NPs; $P> 30 \mathrm{~ms}$) in our Galaxy  can be expressed as a lognormal function \cite{Lorimer2006},
\begin{equation}\label{NPperiod}
f_{\mathrm{dis}}(P) \propto \exp \left[-\frac{(\log P-2.7)^{2}}{2\times (-0.34)^{2}}\right] \,.
\end{equation}
Here, $\log$ is the base 10 logarithm. 
In addition, based on the current observations of about 20 DNSs \cite{Andrews2019,AMM2021}, we assumed that the distribution of the spin periods for the millisecond pulsars (MSPs; $P\leq 30 \mathrm{~ms}$) in the DNS systems follows the uniform distribution 
\begin{equation}\label{MSPperiod}
P \sim \mathcal{U}(10,30)~\rm{ms} \,.
\end{equation}

The probability distribution of pulsar's pseudo luminosity at 1400 MHz $L_{1400}$ can be fitted as follows \cite{FK2006}: 
\begin{equation}\label{eq:fkluminosity}
f_{\mathrm{dis}}(L_{1400}) \propto \exp \left[-\frac{\left(\log L_{1400} + 1.1 \right)^{2}}{2\times 0.9^{2}}\right] \,.
\end{equation}
Note that this formula ignores geometric and beaming factors of the radio beam, thus $L_{1400}$ is not the true physical luminosity of the radio emission from the pulsar \cite{Arzoumanian2002}. On the other hand, although Eq.~(\ref{eq:fkluminosity}) is obtained from the data of NPs, it has been shown that this relationship also holds for MSPs \cite{Bagchi2011}. Thus, in the following simulations, we apply this luminosity distribution to both NPs and MSPs in the DNS systems. 

Given the pulsar distance $D$, the corresponding flux density around 1400~MHz on the Earth is 
\begin{equation}\label{eq:flux}
S_{1400} = \frac{L_{1400}}{D^2} \,.
\end{equation}

\subsection{SNR of integrated pulse profile}
%\subsection{Pulsar simulation and result}
\label{pulsarresult}

According to the standard evolution scenario of the DNSs formed in isolated binaries, their progenitor systems have experienced multiple stages of mass transfer and two supernova explosions (e.g.,~\cite{Tauris2017,sl2018}). It is believed that the first-born NS can be recycled to be an MSP during mass-transfer stages, while the second-born NS is expected to appear as an NP. These are consistent with observed features of pulsars in Galactic DNSs. Thus we assume that each DNS in our simulation is composed of one NP and one MSP. Their spin periods can be assigned according to the distributions given in Eq.~(\ref{NPperiod}) and Eq.~(\ref{MSPperiod}), respectively. The angular radius of the radio beam $\rho$ can be calculated from Eq.~(\ref{rhoperiod}) and Eq.~(\ref{rhoperiod+p}). 
The flux density can be drawn from Eq.~(\ref{eq:fkluminosity}) and Eq.~(\ref{eq:flux}).  

Using the radiometer equation \cite{2004hpa..book.....L}, the SNR of the integrated pulse profile of a pulsar can be calculated by \cite{Lorimer2006}: 
\begin{equation}\label{radiosnr}
\mathrm{SNR} = \frac{S_{1400} \mathcal{G}\sqrt{n_{\mathrm{p}} \Delta \nu \tau}}{\beta T} \sqrt{\frac{P-W_{\mathrm{eff}}}{W_{\mathrm{eff}}}}  \,,
\end{equation}
where $S_{1400}$ is in mJy, $\mathcal{G}$ is the effective telescope gain (K/Jy), $n_{\mathrm{p}}$ is the number of polarizations, $\Delta \nu$ is the observation bandwidth (MHz), $\tau$ is the integration time (s), $\beta$ is the SNR loss factor, $T$ is the system temperature (K), and $P$ is the spin period (ms). The effective pulse width $W_{\mathrm{eff}}$ (ms) can be obtained by 
\begin{equation} \label{effwidth}
W_{\text {eff }}=\sqrt{W^{2}+t_{\text {sam }}^{2}+\Delta t^{2}+\tau_{\mathrm{s}}^{2}} \,,
\end{equation}
where $t_{\text {sam}}$ is the data sampling time in the telescope backend system. The pulse profile broadening in a single filter channel caused by interstellar medium dispersion $\Delta t$ and scattering $\tau_{\mathrm{s}}$ can be evaluated as follows \cite{Bates2014,Bhat2004}
\begin{eqnarray}
\frac{\Delta t}{\mathrm{ms}} &=& ~8.3 \times 10^{6} \times \mathrm{DM} \times \frac{\Delta f_{\mathrm{MHz}}}{f_{\mathrm{MHz}}^{3}}  \,, \\
\log \left(\frac{\tau_{\mathrm{s}}}{\mathrm{ms}}\right) &=& -6.46+0.154 \log (\mathrm{DM}) \nonumber \\
&& +1.07(\log \mathrm{DM})^{2}-3.86 \log f_{\mathrm{GHz}}  \,.
\end{eqnarray}
Here, $\Delta f_{\mathrm{MHz}}$ is the frequency channel width (MHz), $f_{\mathrm{MHz}} = 1400~{\mathrm{MHz}}$, and $f_{\mathrm{GHz} } = 1.4~{\mathrm{GHz}}$. DM is the dispersion measure ($\mathrm{pc/cm^3}$) for which we adopt the YMW16 electron density model \cite{YMW16}.

We set the SNR threshold of the integrated pulse profile to be 9 as in \cite{Bates2014}, above which the pulsar is regarded as detectable. Meanwhile, as one can see from Eq.~(\ref{radiosnr}), the SNR is also determined by the technical specifications of radio telescopes. In order to see the potential to detect the pulsars in our simulations, in the following we consider mock searches of these pulsars based on the Parkes~\cite{Manchester2001}, FAST~\cite{Smits2009,HuangFAST2020}, SKA1-Mid~\cite{2022SKA1-Mid,Marek2020SKA1para}, and SKA-Mid~\cite{Smits2009SKA}, of which the necessary technical specifications and observation parameters are listed in Table~\ref{tab:teleparameter}. Note that the declination ranges of visible sky are $[-90^{\circ},25^{\circ}]$ for Parkes \cite{Parkes_Dec}, $[-14.4^{\circ},65.6^{\circ}]$ for FAST \cite{FAST_Dec}, and $[-90^{\circ},45^{\circ}]$ for SKA1-Mid and SKA-Mid \cite{SKA_Dec}. 
\begin{table}[htb]
    \centering
    \caption{The technical specifications and observation parameters for the pulsar searches based on the Parkes, FAST, SKA1-Mid, and SKA-Mid. The SKA-Mid is assumed to have the same parameters as the SKA1-Mid except for $\mathcal{G}=110~\mathrm{K Jy^{-1}}$. }
    \label{tab:teleparameter}
    \begin{ruledtabular}
    \begin{tabular}{cccc}

     & {Parkes} & {FAST} & {SKA1-Mid} \\ \hline
    Correction factor ~$\beta$ & 1.2 & 1.2 & 1.2 \\
    Gain ~$\mathcal{G}~(\mathrm{K Jy^{-1}})$ & 0.7 & 16.5 & 15 \\
    Integration time~$\tau~(\mathrm{s})$ & 2100 & 600 & 2100 \\
    Sampling time ~$t_{\mathrm{sam}}~(\mathrm{\mu s})$ & 250 & 50 & 64 \\
    System temperature ~$T~(\mathrm{K})$ & 25 & 25 & 30 \\
    Bandwidth~$\Delta \nu~(\mathrm{MHz})$ & 288 & 400 & 300 \\
    Channel width~$\Delta f~(\mathrm{MHz})$ & 3 & 0.39 & 0.02 \\
    Number of polarizations~$n_\mathrm{p}$ & 2 & 2 & 2 \\
    %SNR Threshold & 9 & 9 & 9 \\
    % Declination range &$(-90^{\circ},25^{\circ})$  &$(-14.4^{\circ},65.6^{\circ})$  &$(-90^{\circ},45^{\circ})$  \\

    \end{tabular}
    \end{ruledtabular}
\end{table}

\subsection{Number of detectable pulsar binaries}\label{sec:num}

\begin{figure}[!htbp]%
  \centering
  \subfigure[TQ+Parkes]{\label{fig:PSBinaryNumber_TQ+PMPS}
  \includegraphics[scale=0.4]{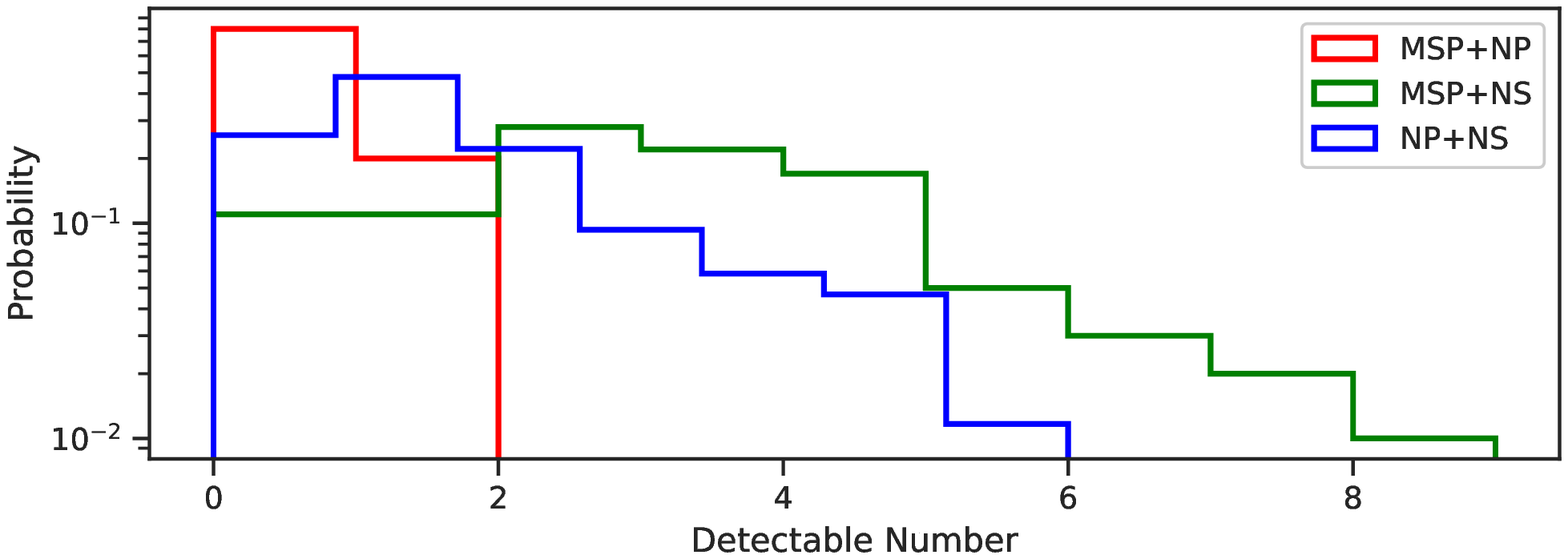}}% 

  \subfigure[TQ+FAST]{\label{fig:PSBinaryNumber_TQ+FAST}
  \includegraphics[scale=0.4]{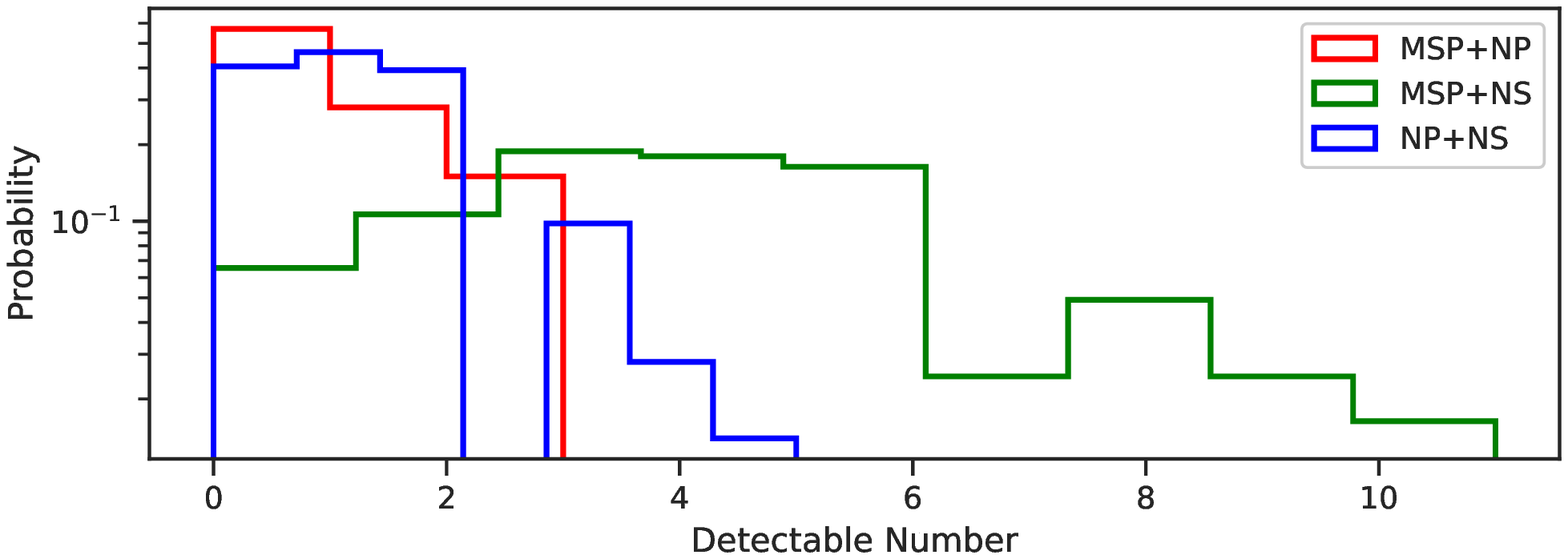}}% 

  \subfigure[TQ+SKA1-Mid]{\label{fig:PSBinaryNumber_TQ+SKA1}
  \includegraphics[scale=0.4]{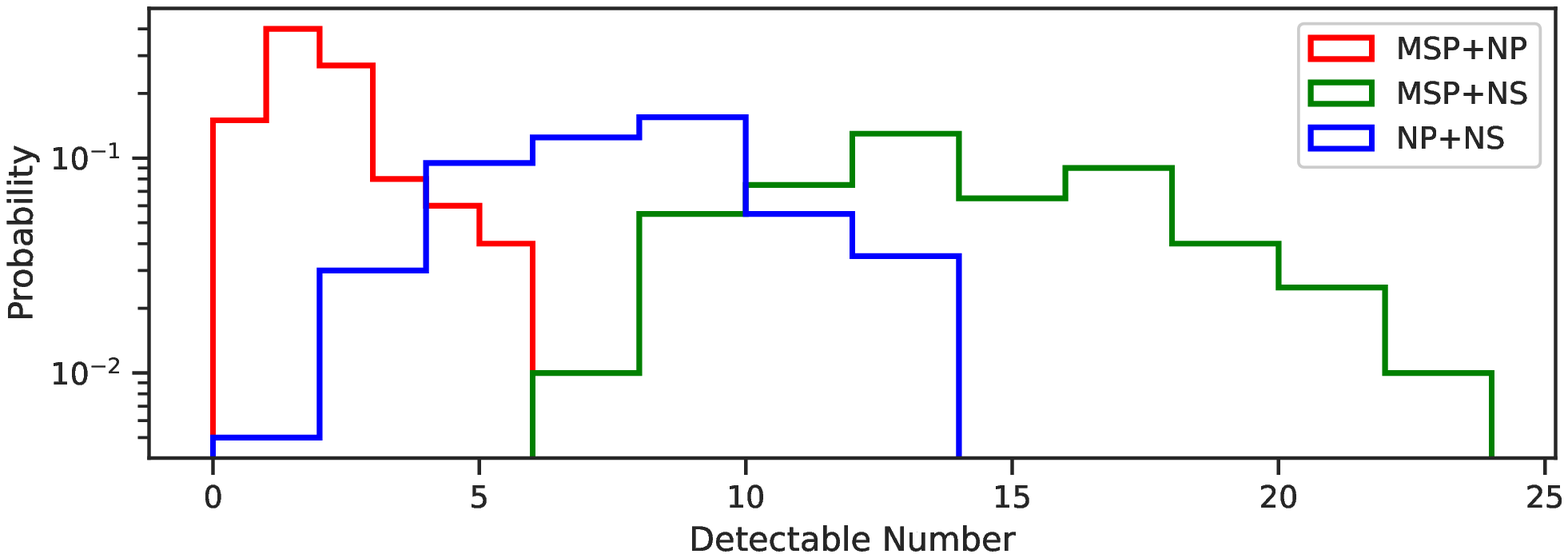}}% 

  \subfigure[TQ+SKA-Mid]{\label{fig:PSBinaryNumber_TQ+fSKA}
  \includegraphics[scale=0.4]{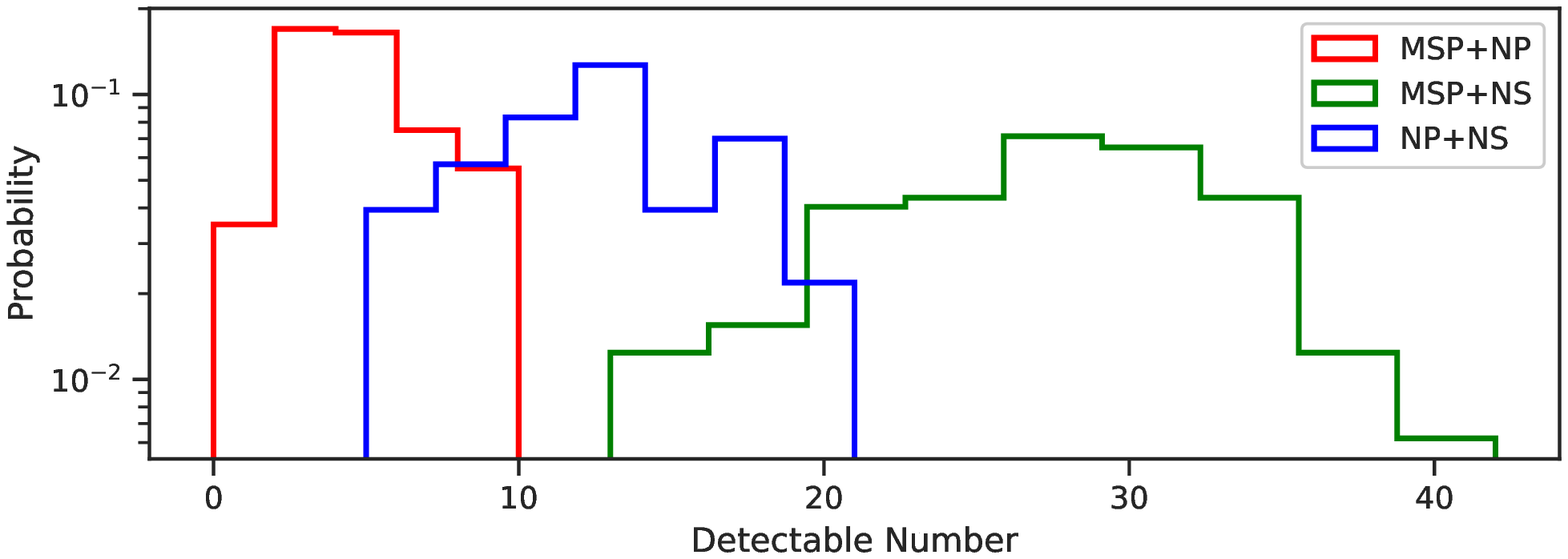}}% 
  \caption{Histograms of the numbers of the pulsar binaries jointly detectable by TQ and the radio telescopes (Parkes, FAST, SKA1-Mid, and SKA-Mid). MSP+NS represents the DNSs that contain an MSP and a NS with no visible radio pulses. }
  \label{fig:GW_Radio_number_TQ+Telescope}
\end{figure}

\begin{figure}[!htbp]%
  \centering
  \subfigure[LISA+Parkes]{\label{fig:PSBinaryNumber_LISA+PMPS}
  \includegraphics[scale=0.4]{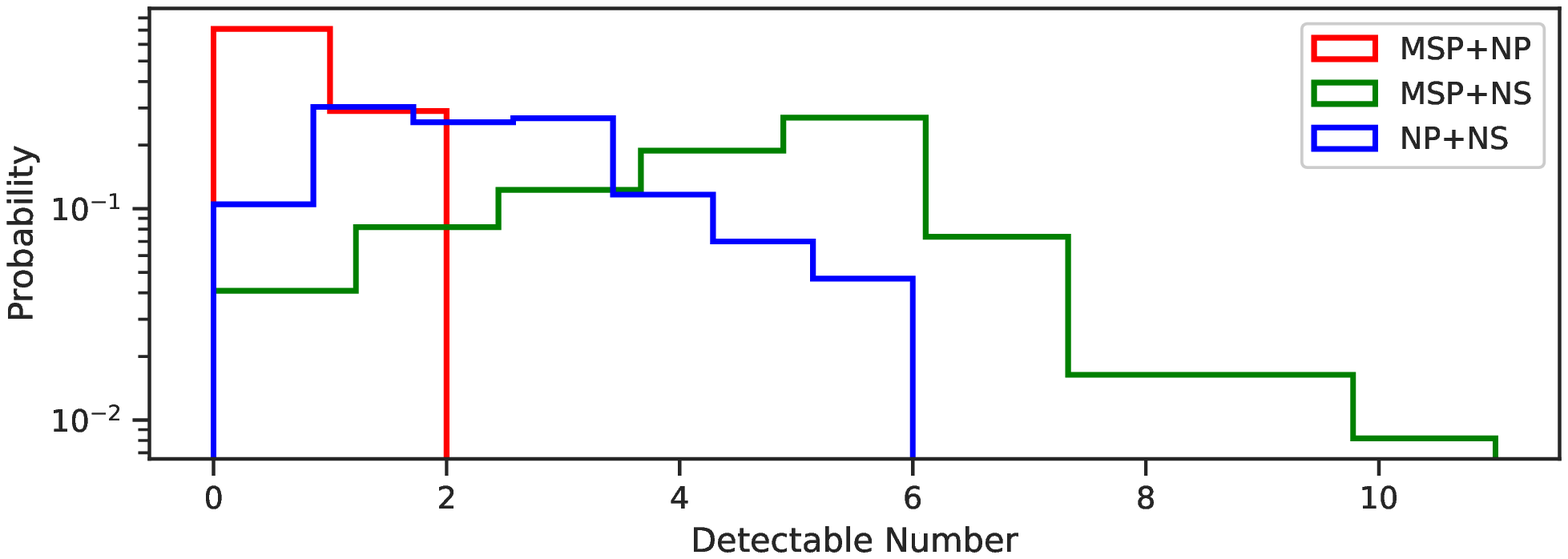}}%

  \subfigure[LISA+FAST]{\label{fig:PSBinaryNumber_LISA+FAST}
  \includegraphics[scale=0.4]{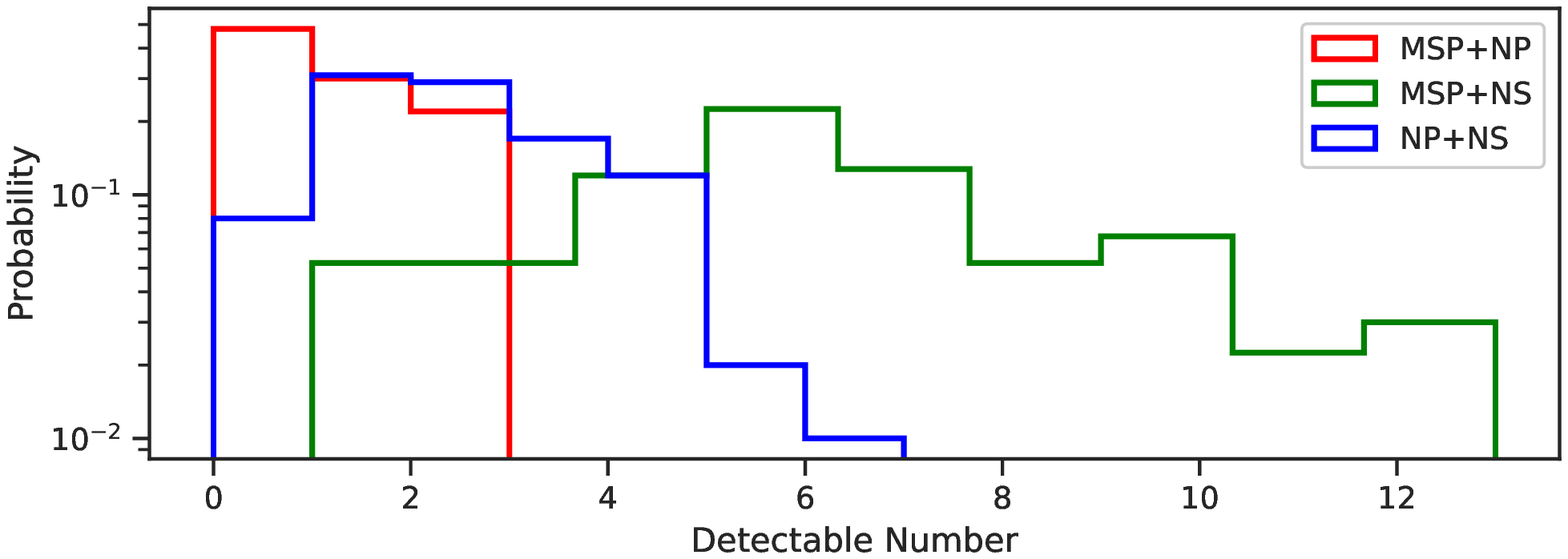}}%

  \subfigure[LISA+SKA1-Mid]{\label{fig:PSBinaryNumber_LISA+SKA1}
  \includegraphics[scale=0.4]{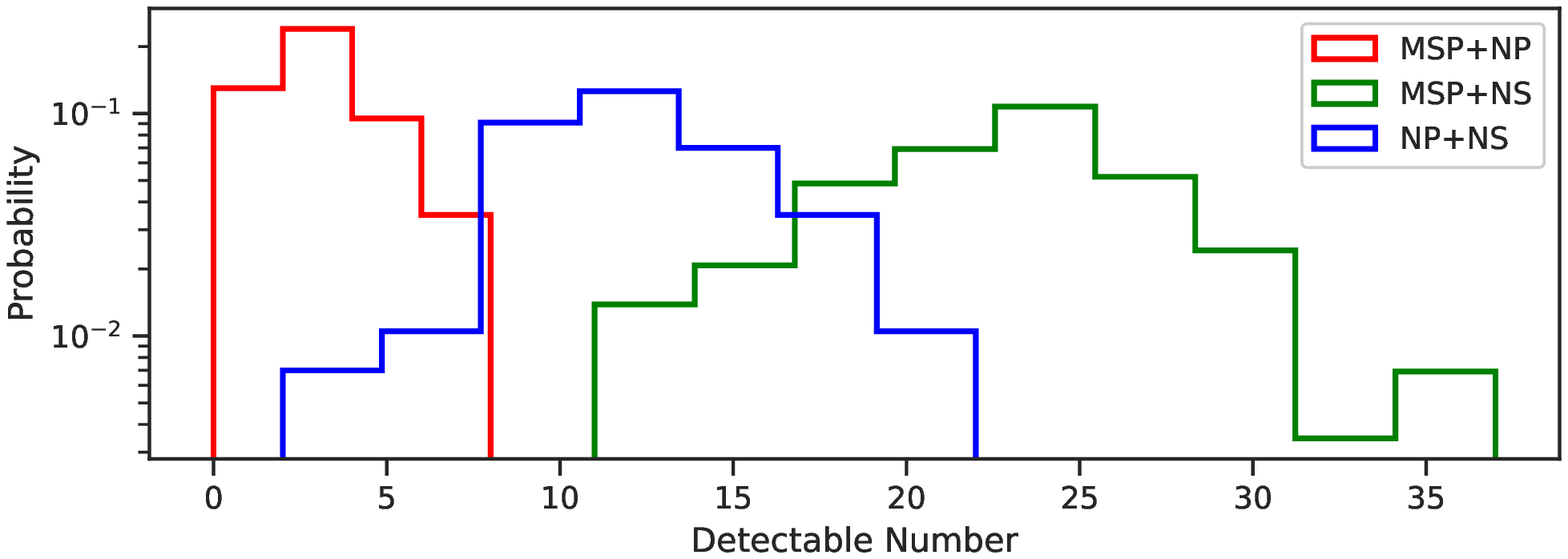}}%

  \subfigure[LISA+SKA-Mid]{\label{fig:PSBinaryNumber_LISA+fSKA}
  \includegraphics[scale=0.4]{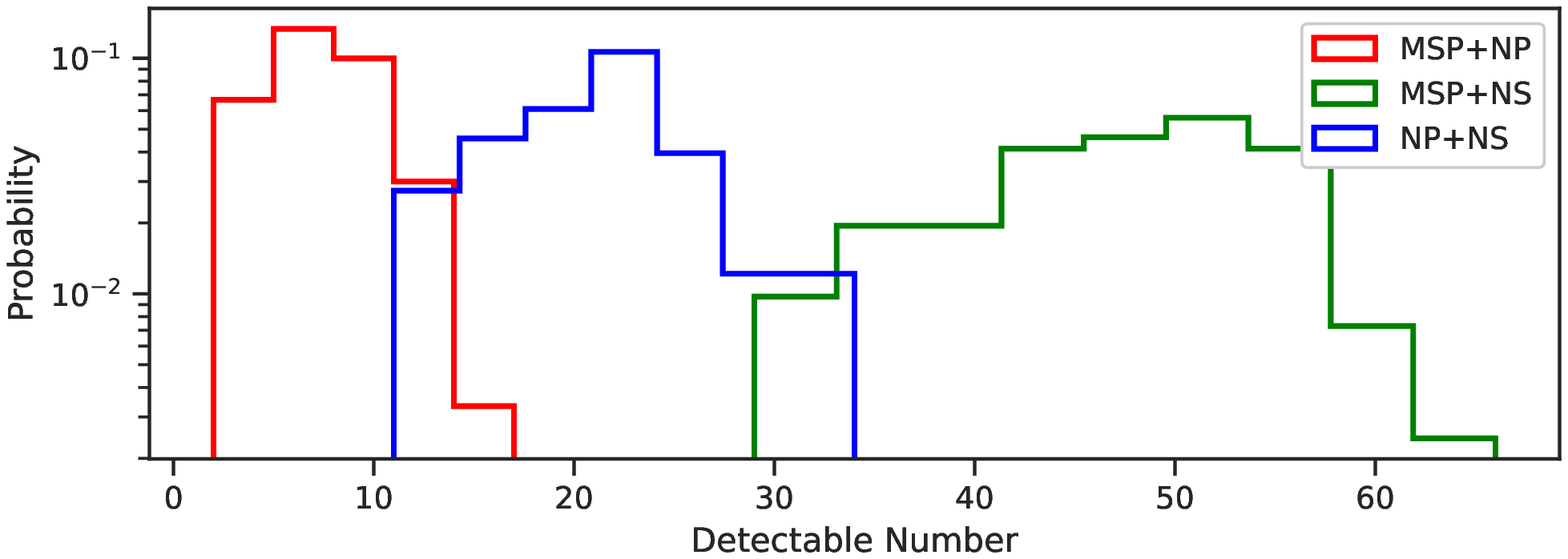}}%
  \caption{Histograms as in Fig. \ref{fig:GW_Radio_number_TQ+Telescope}, but for LISA. }
  \label{fig:GW_Radio_number_LISA+Telescope}
\end{figure}

\begin{figure}[!htbp]%
  \centering
  \subfigure[TQ+LISA+Parkes]{\label{fig:PSBinaryNumber_TQ+LISA+PMPS}
  \includegraphics[scale=0.4]{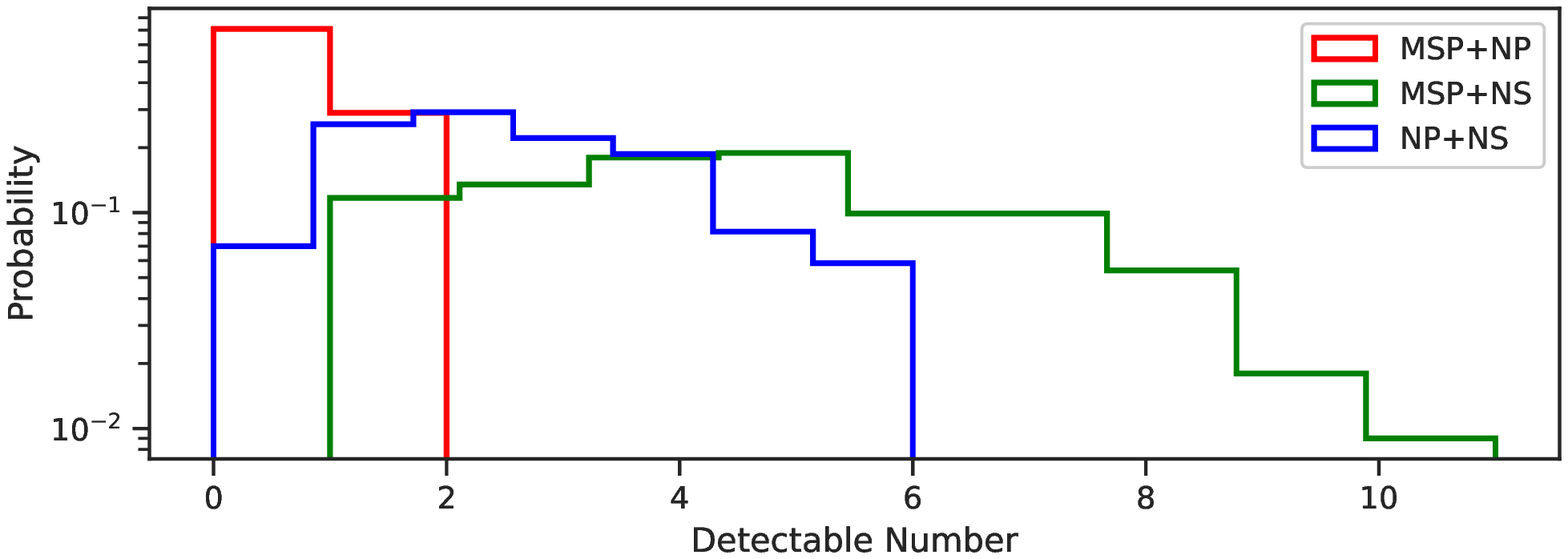}}%

  \subfigure[TQ+LISA+FAST]{\label{fig:PSBinaryNumber_TQ+LISA+FAST}
  \includegraphics[scale=0.4]{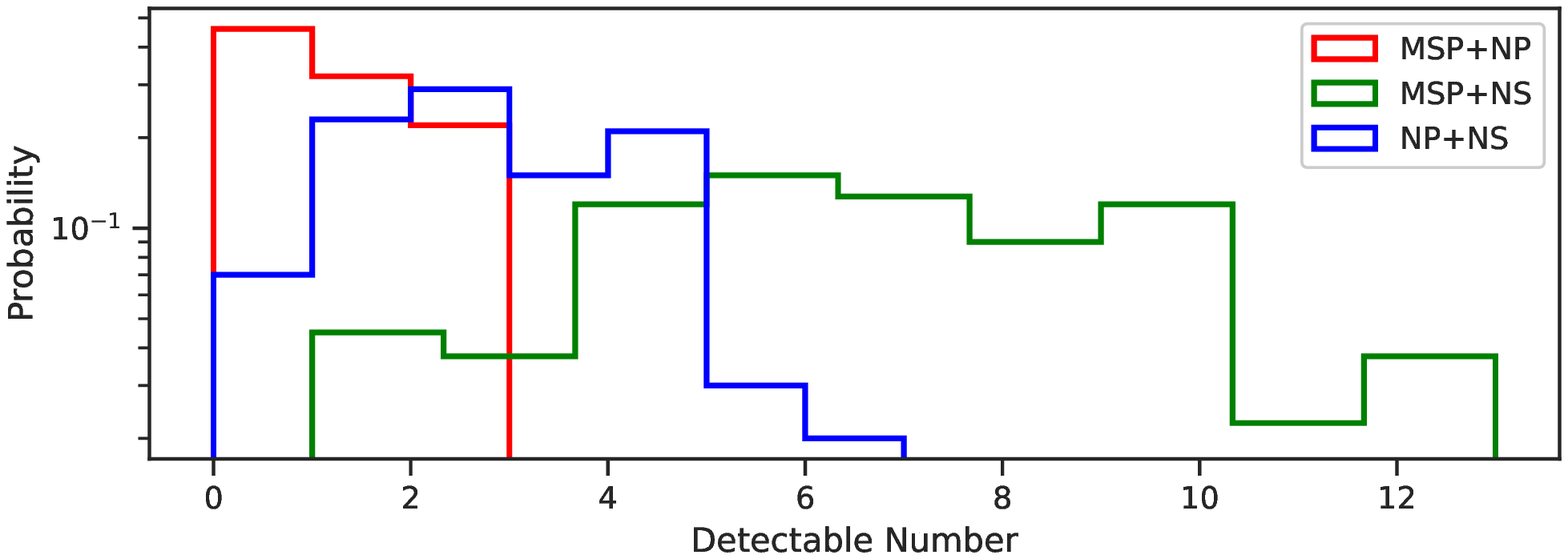}}%

  \subfigure[TQ+LISA+SKA1-Mid]{\label{fig:PSBinaryNumber_TQ+LISA+SKA1}
  \includegraphics[scale=0.4]{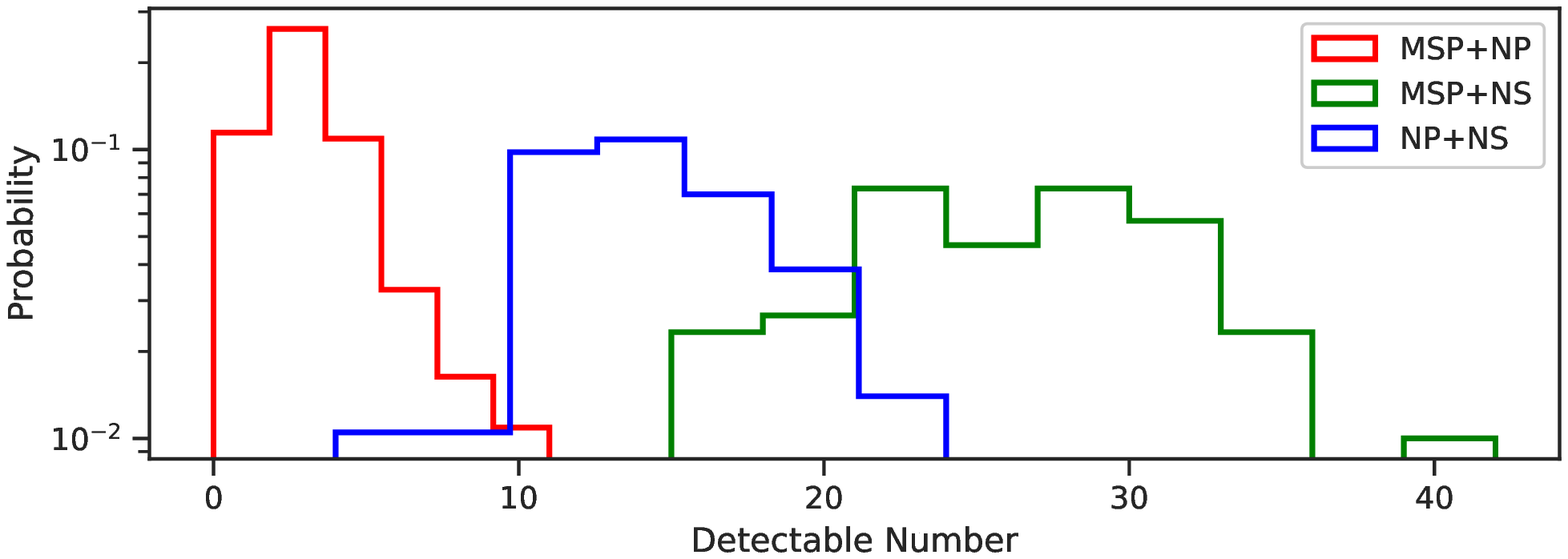}}%

  \subfigure[TQ+LISA+SKA-Mid]{\label{fig:PSBinaryNumber_TQ+LISA+fSKA}
  \includegraphics[scale=0.4]{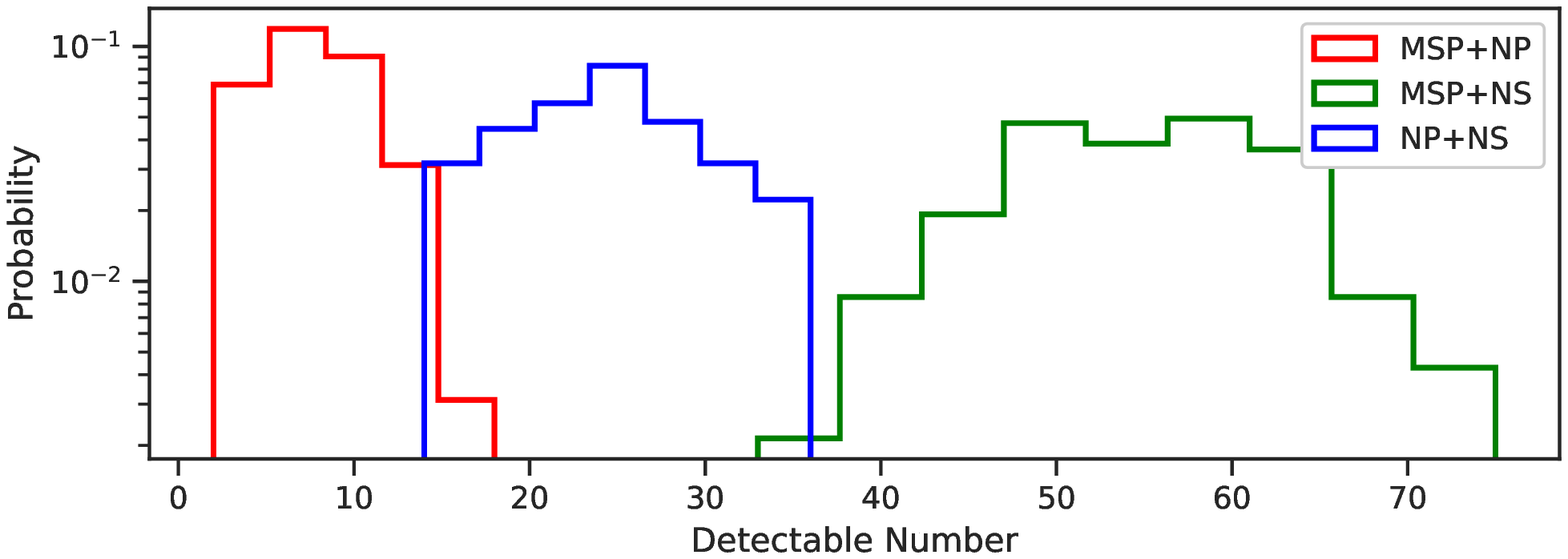}}%
  \caption{Histograms as in Fig. \ref{fig:GW_Radio_number_TQ+Telescope}, but for TQ+LISA. }
  \label{fig:GW_Radio_number_TQ+LISA+Telescope}
\end{figure}

We obtain the numbers of the DNSs in Sec.~\ref{sec:GWresults} that are jointly detectable by the space-borne GW detectors and the radio telescopes. 
%containing pulsars that can be found in the mock surveys based on the Parkes, FAST, SKA1-Mid, and SKA-Mid, respectively. 
The histograms of the numbers from 100 independent Monte Carlo simulations are shown in Fig.~\ref{fig:GW_Radio_number_TQ+Telescope} (for TQ), Fig.~\ref{fig:GW_Radio_number_LISA+Telescope} (for LISA), and Fig.~\ref{fig:GW_Radio_number_TQ+LISA+Telescope} (for TQ+LISA). The corresponding mean values are listed in Table~\ref{tab:GW_pulsarsimulation}.

\begin{table}[htbp]
    \centering
    \caption{The mean numbers of the pulsar binaries from the 100 Monte Carlo simulations 
    that can be jointly detectable by the GW detectors and the radio telescopes with the specifications in Table~\ref{tab:teleparameter}. In each parenthesis, the numbers from left to right are for TQ, LISA, TQ+LISA, respectively.}
    \label{tab:GW_pulsarsimulation}
    \begin{ruledtabular}
    \begin{tabular}{cccc}

             & {MSP-NP}        & {MSP-NS}        & {NP-NS}   \\ \hline
    Parkes   & (0.2, 0.3, 0.3) & (2.7, 4.5, 4.7) & (1.5, 2.3, 2.6) \\
    FAST     & (0.6, 0.8, 0.8) & (4.1, 6.0, 6.6) & (1.2, 2.1, 2.4) \\
    SKA1-Mid & (1.6, 2.7, 3.0) & (14, 23, 26)    & (7.3, 12, 14)   \\
    SKA-Mid  & (4.4, 7.1, 8.0) & (28, 47, 55)    & (13, 21, 24)    \\

    \end{tabular}
    \end{ruledtabular}
\end{table}

Since LISA (red curve in Fig.~\ref{fig:DNSchara}) is more sensitive than TQ (blue curve) around $0.1-10~{\rm mHz}$, the mean numbers of jointly detectable pulsar binaries by LISA and each radio telescope are larger than the corresponding ones for TQ by $50-70\%$ in most of the cases. When operating TQ and LISA together (TQ+LISA), these numbers will be increased by up to $17\%$ than LISA alone. 

In a similar fashion, the numbers of observed pulsar binaries increase with improved sensitivity of the pulsar searches based on the Parkes, FAST, SKA1-Mid, and SKA-Mid. Not surprisingly, due to lower sensitivity, the possibility of detecting short-period double pulsars with an MSP and a NP by the Parkes is quite low; While FAST and SKA1-Mid will have good chances and SKA-Mid will guarantee the detection of several double pulsars. 
All radio telescopes have the potential to detect, at least, several MSP-NS and NP-NS systems jointly with the GW detectors. 
%Especially, SKA1-Mid and SKA-Mid will be likely to detect a few dozen of these systems, which ... \putaref.
The mean numbers for MSP-NS are larger than the ones for NP-NS binaries in all cases, this is due to the fact that the angular radius of the radio beam (cf., Eq.~(\ref{rhoperiod})) for MSP is typically larger than NP. 
FAST is expected to discover 5-10 pulsar binaries detected by TQ, LISA, and TQ+LISA; While SKA1-Mid (SKA-Mid) will further increase the numbers for FAST by a factor of 4 (8).

\subsection{Distances of detectable pulsar binaries}\label{sec:distance}

\begin{figure}[!htbp]
\centering
\includegraphics[scale=0.42]{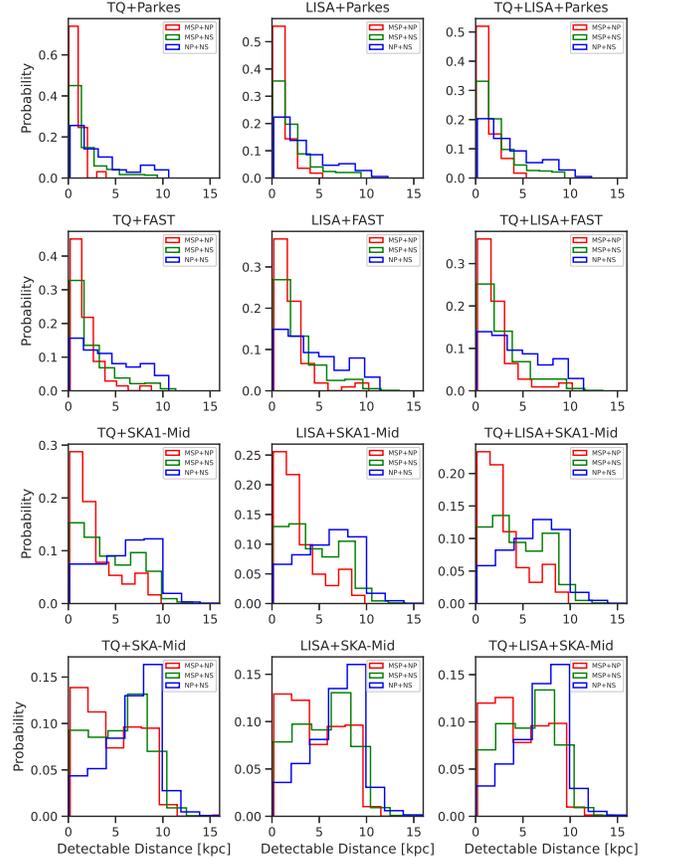}
\caption{ Histograms of the distances for the pulsars in Sec.~\ref{sec:num} that are simultaneously detectable by the GW detectors and the radio telescopes. }
\label{fig:Distance_GW_Radio}
\end{figure}

For those DNSs in Sec.~\ref{sec:GWresults} that are jointly detectable by the space-borne GW detectors and the radio telescopes, the histograms of their distances from 100 independent Monte Carlo simulations are shown in Fig.~\ref{fig:Distance_GW_Radio}, the corresponding mean values are listed in Table~\ref{tab:GW_radio_distance}. 

The average distances of pulsar binaries increase with the increasing sensitivity of the radio telescopes. MSP-NP binaries tend to be in the closer region of 1-5 kpc, while NP-NS (MSP-NS) binaries are in the range of 3-7 kpc (2-6 kpc). 

\begin{table}[htbp]
    \centering
    \caption{ The mean distances (in unit of kpc) of the pulsar binaries from the 100 Monte Carlo simulations 
    that can be jointly detectable by the GW detectors and the radio telescopes. In each parenthesis, the values from left to right are for TQ, LISA, TQ+LISA, respectively.} 
    \label{tab:GW_radio_distance}
    \begin{ruledtabular}
    \begin{tabular}{cccc}

             & {MSP-NP}        & {MSP-NS}        & {NP-NS}   \\ \hline
    Parkes   & (0.9, 1.1, 1.2) & (1.8, 2.1, 2.2) & (3.4, 3.5, 3.6) \\
    FAST     & (1.7, 2.0, 2.1) & (2.5, 2.7, 2.8) & (4.4, 4.4, 4.5) \\
    SKA1-Mid & (2.8, 2.8, 3.0) & (4.3, 4.4, 4.5)    & (5.8, 5.8, 5.9)   \\
    SKA-Mid  & (4.6, 4.6, 4.7) & (5.4, 5.5, 5.6)    & (6.6, 6.7, 6.8)    \\

    \end{tabular}
    \end{ruledtabular}
\end{table}

\subsection{Timing accuracies of detectable pulsars}

The total rms error of TOA measurement ($\sigma_{\mathrm{t}}$) is mainly determined by the radiometer noise ($\sigma_{\mathrm{r}}$) and the pulse phase jitter noise ($\sigma_{\mathrm{j}}$) \cite{Cordes2010, Wang2015},
\begin{equation}\label{eq:rms}
\sigma_{\mathrm{t}} = \sqrt{\sigma_{\mathrm{j}}^2+\sigma_{\mathrm{r}}^2} \,.
\end{equation}
Here, 
\begin{equation}\label{eq:jit}
\sigma_{\mathrm{j}} \approx  0.28 \, W_{\mathrm{eff}} \sqrt{\frac{P}{\tau}} \,,
\end{equation}
and %
\begin{equation}\label{eq:rad}
\sigma_{\mathrm{r}} \approx  \frac{W_{\mathrm{eff}} T}{S_{1400} \mathcal{G} \sqrt{2 \Delta \nu \tau}} \sqrt{\frac{W_{\mathrm{eff}}}{P-W_{\mathrm{eff}}}} \,. 
\end{equation}

\begin{figure}[!htbp]
\centering
\includegraphics[scale=0.4]{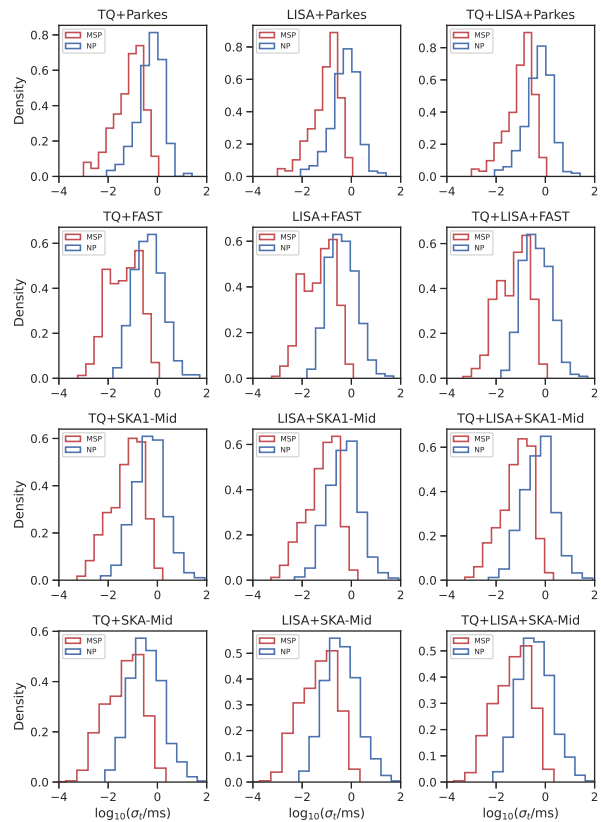}
\caption{Histograms of the total rms errors of the TOA measurements for the pulsars in Sec.~\ref{sec:num} that are 
simultaneously detectable by the GW detectors and the radio telescopes. In each panel, the pulsars from 100 Monte-Carlo simulations are combined. The red and blue histograms represent the results for the MSPs and NPs, respectively. }
\label{fig:GW_Radio_sigma_t}
\end{figure}

Based on the equations above, we calculate $\sigma_{\mathrm{t}}$ for the detectable MSPs and NPs, respectively, combined from 100 Monte-Carlo simulations in Sec.~\ref{sec:num}. Each column of the histograms in Fig.~\ref{fig:GW_Radio_sigma_t} shows the $\sigma_{\mathrm{t}}$ for the corresponding cases in Figs.~\ref{fig:GW_Radio_number_TQ+Telescope}, \ref{fig:GW_Radio_number_LISA+Telescope}, and  \ref{fig:GW_Radio_number_TQ+LISA+Telescope}. 
%
% shows the distributions of the rms errors of TOA measurements for the pulsars that can be detected both by the GW detectors and radio telescopes. As in Sec.~\ref{sec:num}, there are 100 independent Monte Carlo simulations. 
As one can see, the most probable values of the total rms error distributions for the MSPs are a factor of $\sim 1.3-8.2$ smaller than that for the NPs. This is mainly due to the narrower pulse widths of the MSPs than that of the NPs. 
In addition, for SKA-Mid with the highest sensitivity, the total rms errors for the MSPs can be as low as $70~\mathrm{ns}$ with most ones concentrated around $100~\mathrm{\mu s}$; while $\sigma_{\mathrm{t}}$ for NPs can be as low as $5~\mathrm{\mu s}$ with most ones concentrated around  $130~\mathrm{\mu s}$. 

\section{Conclusions}\label{sec:conclusion}

In this work, we simulate the DNS population in the Galactic disk that will merge within the next $10~{\rm Myr}$. The DNS merger rate is adopted from the recent LIGO results. The numbers of the detectable DNSs and the associated parameter estimation accuracies obtained by TQ, LISA, and TQ+LISA are estimated by FIM in 100 Monte Carlo simulations. 
Based on the beam geometry and empirical distributions of pulsar spin period and luminosity, we get the numbers and distances of these DNSs comprising radio pulsars that can be potentially detected by the Parkes, FAST, SKA1-Mid, and SKA-Mid. In addition, we estimate the timing accuracies of the TOAs for these pulsars by evaluating their dominating radiometer and jitter noises. 

There are caveats caused by the assumptions used in our simulations.
Firstly, we follow \cite{Andrews2020} in setting the orbital parameters of B1913+16 at the merger as the benchmark ($P_\mathrm{b0} = 0.2 ~\mathrm{s}$, $e_0 = 5 \times 10^{-6}$) in the calculation of the orbital parameters of the simulated DNSs. However, when we set J0737-3039 \cite{Kramer2006Sci} as the benchmark, i.e., $P_\mathrm{b0} = 0.18 ~\mathrm{s}$ and $e_0 = 1 \times 10^{-6}$ at the merger, the eccentricities of the simulated DNSs are generally smaller ($e \lesssim 0.04$) than the ones in Sec.~\ref{sec:GWdetection}, according to the orbital evolution equations in Sec.~\ref{sec:DNSpopulation}. By repeating the simulation procedure in Sec.~\ref{sec:GWdetection}, we find that the distributions of the estimated parameters for the latter case remain unchanged  except for $\Delta e/e$, the overall amplitude of which increases by a factor of 5 while the power law form in Eq. (\ref{Deltae2e}) holds the same.
Secondly, there are only five MSPs in DNSs are discovered to date \cite{Andrews2019}, due to the lack of prior knowledge, we assume that the distribution of their spin periods obeys a uniform distribution (cf. Eq.~\ref{MSPperiod}). 
This will be improved when more MSPs are discovered in future radio pulsar surveys. 
We also assume that the luminosity distribution of the pulsars in tight binaries is the same as that of the normal pulsars. If a long period of time has elapsed from the wide binaries (at the beginning of their formation) to the present tight binaries, the luminosity of pulsars in the latter population may be dimmer because of the decay of the luminosity with time~\cite{Bates2014}.

In this work, we only consider the SKA1-Mid and SKA-Mid in pulsar observations. 
However, the SKA1-Low and SKA-Low and their precursors or pathfinders working mainly in $50-350$~MHz may have an important contribution in the search for pulsars in DNSs, especially when a pulsar has a steep power spectrum in flux density \cite{2021ApJ...911L..26S}. Furthermore, observation of pulsars in low frequencies can effectively improve the timing accuracy through the measurement and mitigation of the dispersion effects, and for the pulsars near the ecliptic plane, the solar wind effects \cite{2022MNRAS.511.3937K}. 

For the Galactic DNSs, both LISA and TQ have the potential to detect their 3rd harmonic signals (cf. Eq. (\ref{eq:hplushcross})). Combining with the 2nd harmonic signals (cf. Fig.~\ref{fig:DNSchara}), we can determine the total masses of these eccentric DNSs through the GW frequency shift caused by the periastron advance \cite{Seto2001}. Thus, we can obtain the component masses of the binary by their chirp mass and total mass.  
With the aid of multi-messenger observation realized by space-borne GW detectors and radio telescopes, these ultra-compact binary systems with short orbital periods ($\lesssim 10$~min) are expected to probe the highly relativistic regime of astrophysics, such as constrain the equation of state of NS with remarkable fidelity \cite{Thrane2020} or tight constraint on the parameter space for specific alternative theories of gravity \cite{Miao2021}. These considerations are currently under our investigation.

\begin{acknowledgments}
Y.W. gratefully acknowledges support from the National Key Research and Development Program of China (Nos. 2022YFC2205201 and 2020YFC2201400), 
 the National Natural Science Foundation of China (NSFC) under Grants No. 11973024, Major Science and Technology Program of Xinjiang Uygur Autonomous Region (No. 2022A03013-4), and Guangdong Major Project of Basic and Applied Basic Research (Grant No. 2019B030302001). 
J.W.C acknowledges the support from China Postdoctoral Science Foundation under Grant No. 2021M691146. 
We thank the anonymous referees for helpful comments and suggestions. 

\end{acknowledgments}

\bibliography{reference}

\end{document}